\input epsf
\documentstyle{l-aa}
 
\begin{document}
\def\mspc{M_\odot.pc^{-3}}
\def\msol{M_\odot}
\def\mso{M_\odot}
\def\NeG{NextGen}
\def\te{T_{eff}}
\def\rs{r_s}
\def\mh{[M/H]}
\def\lsol{L_\odot}
\def\rsol{R_\odot}
\def\tsol{T_{eff_\odot}}
\def\Zsol{Z_\odot}
\def\wig#1{\mathrel{\hbox{\hbox to 0pt{%
          \lower.5ex\hbox{$\sim$}\hss}\raise.4ex\hbox{$#1$}}}}
\newcommand\etal{{\it et al.}}

%references

\def\aj{AJ}                  % Astronomical Journal
\def\araa{ARA\&A}             % Annual Review of Astron and Astrophys
\def\apj{ApJ}                 % Astrophysical Journal
\def\apjl{ApJ}                % Astrophysical Journal, Letters
\def\apjs{ApJS}               % Astrophysical Journal, Supplement
\def\ao{Appl.Optics}          % Applied Optics
\def\apss{Ap\&SS}             % Astrophysics and Space Science
\def\aap{A\&A}                % Astronomy and Astrophysics
\def\aapr{A\&A~Rev.}          % Astronomy and Astrophysics Reviews
\def\aaps{A\&AS}             % Astronomy and Astrophysics, Supplement\def\azh{AZh}                 % Astronomicheskii Zhurnal
\def\baas{BAAS}               % Bulletin of the AAS
\def\jrasc{JRASC}             % Journal of the RAS of Canada
\def\memras{MmRAS}            % Memoirs of the RAS
\def\mnras{MNRAS}             % Monthly Notices of the RAS
\def\pra{Phys.Rev.A}          % Physical Review A: General Physics
\def\prb{Phys.Rev.B}          % Physical Review B: Solid State
\def\prc{Phys.Rev.C}          % Physical Review C
\def\prd{Phys.Rev.D}          % Physical Review D
\def\prl{Phys.Rev.Lett}       % Physical Review Letters
\def\pasp{PASP}               % Publications of the ASP
\def\pasj{PASJ}               % Publications of the ASJ
\def\qjras{QJRAS}             % Quarterly Journal of the RAS
\def\skytel{S\&T}             % Sky and Telescope
\def\solphys{Solar~Phys.}     % Solar Physics
\def\sovast{Soviet~Ast.}      % Soviet Astronomy
\def\ssr{Space~Sci.Rev.}      % Space Science Reviews
\def\zap{ZAp}                 % Zeitschrift fuer Astrophysik

%\thesaurus{02(02.01.4; 02.04.2; 08.01.1; 08.03.2)}
\thesaurus{}
\title{ Structure and evolution of low-mass stars}

\author{{\sc Gilles Chabrier and Isabelle Baraffe}}

\institute{Centre de Recherche Astrophysique de Lyon (UMR 5574 CNRS)\\
Ecole Normale Sup\'erieure de Lyon, 69364 Lyon Cedex 07, France\\
(chabrier, ibaraffe @cral.ens-lyon.fr)}

%\authoraddr{ Ecole Normale Sup\'erieure de Lyon, 69364 Lyon Cedex 07, France}
%\authoremail{chabrier, ibaraffe, cral.ens-lyon.fr}

\date{Received date ; accepted date}

\maketitle

\markboth{G. Chabrier and I. Baraffe: Structure and evolution 
of low-mass stars} {}

%\begin{center}
%$\underline{Submitted\ to}$: {\sl ApJ Letters}

%\bigskip
%\bigskip
%\underline{Version}: \today
%\end{center}
%\bigskip
%\bigskip

\begin{abstract}
We present extensive calculations of the structure and the evolution of low-mass
stars in the range 0.07-0.8 $\msol$, for metallicities $-2.0\le \mh \le 0.0$.
These calculations are based on the most recent description of the
microphysics characteristic of these dense and cool objects and on the lattest
generation of grainless non-grey atmosphere models. We examine the evolution of the
different mechanical and thermal properties of these objects as a function of
mass and metallicity. We also demonstrate the inaccuracy of grey
models and $T(\tau)$ relationships under these conditions. We provide detailed
tables of the mass-radius-luminosity-effective temperature relations
for various
ages and metallicities, aimed at calibrating existing or future observations
of low-mass stars and massive brown dwarfs.
We derive new hydrogen-burning minimum masses, within the afore-mentioned
metallicity range. These minimum masses are found to be smaller than
previous estimates, a direct consequence of non-grey effects.

At last, we examine the evolution of the abundance of light elements,
$Li, Be$ and $B$, as a function of age, mass and metallicity.

\end{abstract}

\section{Introduction}

Accurate modelling of the mechanical and thermal properties of very-low-mass stars (VLMS), or M-dwarfs, defined hereafter as objects with
masses below $\sim$0.8$\msol$, is of prior importance for a wide range of  physical and astrophysical reasons, from the understanding of fundamental
problems in basic physics
to astrophysical and cosmological implications. VLMS are compact objects, with characteristic radii in the range
$ 0.1\wig < R/R_\odot \wig < 0.7$. Their central densities and temperatures are respectively of the order
of $\rho_c\approx 10-10^3$ g cm$^{-3}$ and $T_c\approx 10^6-10^7$ K, so that correlation effects between the particles dominate the kinetic contribution
in the interior stellar plasma.
Effective temperatures of VLMS are below $T_{eff}\approx 5000$ K, and surface gravities $g=GM_\star/R_\star^2$ are
in the range $\log g\approx 3.5-5.5$. These conditions show convincingly that the modelling of VLMS requires a correct description
of non-ideal effects in the equation of state (EOS) and the nuclear reaction
rates, and a derivation of accurate models for dense and cool atmospheres, where molecular opacity becomes eventually
the main source of absorption. Several ground-based and space-based IR missions
are now probing the VLMS wavelength range ($\lambda \approx 1-10\, \mu$) down to the end of the main-sequence (MS), reaching sometimes
the sub-stellar domain. These existing or future surveys will
produce a substantial wealth of data, stressing the need for accurate theroretical models. Indeed the ultimate goal of VLMS
theory is an accurate calibration of observations, temperature, luminosity
and above all the mass, with the identification of
genuine brown dwarfs. At last, VLMS represent the
major component ($>70\%$) of the stellar population of the Galaxy.
A correct determination of the contribution
of these objects to the Galactic mass budget, both in the central
parts and in the outermost halo, requires the derivation
of reliable mass functions for VLMS, and thus accurate theoretical
{\it mass-luminosity}
relationships for various metallicities.

Tremendous progress has been realized within the past decade in the field of VLMS, both from the observational and
theoretical viewpoints.
 From the theoretical point of view, the most recent
benchmarks in
the theory, without being exhaustive, have been made by D'Antona \& Mazzitelli (1985, 1994), who initiated the research in the field,
the MIT group (Dorman, Nelson \& Chau 1989; Nelson, Rappaport \& Joss 1986, 1993)
and the Tucson group (Lunine et al.
1986; Burrows, Hubbard \& Lunine 1989; Burrows et al. 1993). So far, all these models, however, failed to reproduce the
observations at the bottom of the VLMS sequence, predicting substantially too large temperatures for a given luminosity
(see e.g. Monet et al. 1992). This shortcoming of the theory made a reasonable identification of the observational H-R diagram
elusive. 
%stemed from overestimated stellar radii
Such a discrepancy stemed from incorrect stellar radii (and adiabatic gradients),
a consequence of inaccurate EOS,
but most importantly from the use of {\it grey} atmosphere models. These points will be largely examined in \S2.4 and \S2.5.
A significant breakthrough in the structure and evolution of VLMS
was made recently by the Tucson group, who first derived evolutionary models based on
non-grey atmosphere models (Saumon et al. 1994), although for zero-metallicity, and by the Lyon group (Baraffe et al. 1995, BCAH95; 1997, BCAH97; Chabrier et al. 1996)
who derived evolutionary models based on the Allard-Hauschildt (1995a, AH95; 1997, AH97) non-grey model atmospheres for various finite metallicities.
The BCAH95 models were shown to improve significantly the afore-mentioned discrepancy at the bottom of the MS. These initial calculations
have now been improved substantially. The aim of the present paper is
to present a complete description of the
physics entering the theory of VLMS and to relate
the properties of these objects to well understood physical
grounds. Extensive comparisons with available observations,
color-magnitude diagrams and mass-magnitude relationships,
will be presented in  companion papers (BCAH97;
Allard et al. 1997a).

The present paper is organized as follows. In \S 2, we describe the input physics which enters the
present theory, EOS, enhancement factors of the nuclear reaction rates, atmosphere models and boundary conditions.
Evolutionary models are presented in \S 3, together with the prediction of the abundances of light elements ($^7$Li,$^9$Be,$^{11}$B)
along evolution and the burning minimum masses for these elements.
We also derive a new limit for the {\it hydrogen} burning minimum mass (HBMM), i.e. the
brown dwarf limit, which is found to be lower than previous estimates,
a direct consequence of non-grey effects.
% Special attention will be devoted to deuterium-burning during the PPI %reaction,
%for deuterium burning is found to proceed much faster than %deuterium-mixing, therefore never reaching an equilibrium concentration.
The mass-dependence of photospheric quantities is examined in \S4.
Section 5 is devoted
to the concluding remarks.
%Mass-Luminosity and Mass-spectral type relationships are shown in \S4 and compared with observationally
%determined relations. The application of the models to the two eclipsing binary VLMS CM Draconis and YY Geminorum are
%discussed in \S 5. Section 6 is devoted to the concluding remarks.

\section{Input physics}

The Lyon evolutionary code has been originally developped at the
G\"ottingen Observatory (Langer et al. 1989; Baraffe and El Eid, 1991 and references therein)
and is based on one-dimensional implicit equations of stellar structure,
solved with the Henyey method (Kippenhahn and Weigert, 1990).
Convection is described by the mixing-length theory. Throughout the
present paper, we use a mixing length equal to the pressure scale height, l$_{mix}$ = H$_p$ as a reference
calculation. As discussed in \S 3.2, the choice of this parameter is
inconsequential for the evolution of objects below $\sim 0.6\,\msol$.
For larger masses, the dependence of the results on the mixing length is
examined in BCAH97, along comparison with observations.
The updated Livermore opacities (OPAL,
Iglesias \& Rogers 1996) are used for the inner structure ( T $>$ 10 000K).
The effect of the improved OPAL opacities compared to the previous generation
(Rogers \& Iglesias, 1992) on the evolution of VLM stars
(BCAH95;
Chabrier, Baraffe \& Plez 1996) is found to be
negligible, affecting the effective temperature by less than 1\%
and the luminosity by less than 3\%  for a given mass.
For lower temperatures, we use the Alexander and Fergusson (1994) opacities. The helium fraction in the calculations is Y=0.275 for solar-like
metallicities and Y=0.25 for metal-depleted abundances.

An adequate theory for stellar evolution requires i) an accurate EOS, ii) a correct treatment
of the nuclear reaction rates, iii) accurate
atmosphere models and iv) a correct treatment of the boundary conditions between the interior and the atmophere
profiles along evolution. Each of these inputs is discussed in the following sub-sections.

\subsection{The equation of state}

Interior profiles of VLMS range from $\sim 4000$ K and $\sim 10^{-6}$ g  cm$^{-3}$ at the base of the photosphere (defined as
$R_{ph}=\sqrt{(L/4\pi\sigma\te^4)}$) to
 $\sim 10^7$ K and $\sim 100$ g cm$^{-3}$ at the center for a 0.6 $\msol$ star, and
from $\sim 2800$ K and $\sim 10^{-5}$ g  cm$^{-3}$ to
$\sim 5 \, 10^6$ K and $\sim 500$ g cm$^{-3}$ for a 0.1 $\msol$, for solar metallicity.
Within this temperature/density range, molecular hydrogen and atomic helium are stable in the outermost part of the stellar
envelope, while most of the bulk of the star (more than 90\% in mass) is under the form of a fully ionized H$^+$/He$^{++}$ plasma. Therefore
a correct EOS for VLMS must include a proper treatment not-only of temperature-ionization and dissociation,
well described by the Saha-equations in an ideal gas, but most importantly of {\it pressure}-ionization and
dissociation, as experienced along the internal density/temperature profile, a tremendously more complicated task.
Moreover, under the central conditions of these stars, the fully ionized
hydrogen-helium plasma is characterized by a plasma coupling parameter $\Gamma=(Ze)^2/akT\propto (Z^2/A^{1/3})\, (\rho_6^{1/3}/T_8)\approx 0.5-5$
for the classical ions ($a$ is the mean inter-ionic distance, $A$ is the atomic mass and $\rho$ the {\it mass}-density)
and by a quantum coupling parameter $\rs=<Z>^{-1/3}a/a_0=1.39/(\rho/\mu_e)^{1/3}
\approx 0.1-1$ ($a_0$ is the electronic Bohr radius and $\mu_e^{-1}={<Z>\over <A>}$ the electron
mean molecular weigth)
for the degenerate electrons. These parameters show that both the ions and the electrons are
strongly correlated. A third characteristic parameter is the so-called {\it degeneracy parameter} 
$\psi=kT/kT_F\approx 3\times 10^{-6}\, T\,  (\mu_e/\rho)^{2/3}$,
where $kT_F$ is the electron Fermi energy. The
classical (Maxwell-Boltzman) limit corresponds to $\psi \rightarrow + \infty$, whereas $\psi \rightarrow 0$ corresponds
to complete degeneracy. The afore-mentioned thermodynamic conditions yield $\psi \approx 2-0.1$ in the interior of VLMS along the characteristic mass range, 
%*** ULTRA VERIFIER ***
implying that {\it finite-temperature} effects must be included to describe accurately the thermodynamic properties
of the correlated electron gas. At last, the Thomas-Fermi wavelength $\lambda_{TF}=
\bigl( kT_F/(6\pi n_e e^2)\bigr)^{1/2}$ (where $n_e$ denotes the electron particle density) is of the
order of the mean inter-ionic distance $a$, so that the electron gas is {\it polarized} by the ionic field, and
electron-ion coupling must be taken into account in 
the plasma hamiltonian. Such a detailed treatment of strongly
correlated, polarisable classical and quantum plasmas, plus an accurate description of pressure partial ionization
represent a severe challenge for theorists. Several steps towards the derivation of such an accurate EOS for VLMS
have been done since the pioneering work of Salpeter (1961) and we refer
the reader to Saumon (1994) and Saumon, Chabrier and VanHorn (1995) for
a review and a comparison of the different existing EOS for VLMS.
%Marley and Hubbard have derived an EOS which is restricted
%to the domain of extremely low-mass stars ($M<0.2,\msol$), brown dwarfs %and planets.
%But the most extensive work, and the most widely used EOS within the VLMS %has been the so-called FGVH EOS (Fontaine,
%Graboske \& VanHorn 1977) for the past 15 years or so.
In the present calculations, we use the Saumon-Chabrier (SC) EOS
(Saumon 1990; Chabrier 1990; Saumon \& Chabrier 1991, 1992; Saumon, Chabrier \& VanHorn 1995; SCVH), specially
devoted to the
description of low-mass stars, brown dwarfs and giant planets. This EOS
presents a consistent treatment of pressure ionization and includes
significant improvements w.r.t. previous calculations in the treatment of
the correlations in dense plasmas. This EOS is tied to
available Monte Carlo simulations and high-pressure shock-wave experimental data (see SCVH and references therein for details).
%the SCVH EOS does include experimentally derived potentials and
%compares well with available high-pressure shock-wave experiments in the %molecular domain, and with Monte-Carlo simulations
%in the fully-ionized, metallic domain.
%A detailed description of this EOS, and extensive comparison with the %afore-mentioned EOS, has been
%presented in SCVH and in Saumon (1994) and we send the reader to these %references for more information.
As shown in Saumon (1994) and SCVH,
significant differences exist between the SC EOS and other
VLMS EOS, for the pressure-density relations and for the adiabatic gradients,
so that we expect substantial differences in the derived stellar radii and entropy profiles.
%***** A severe test for the EOS of VLMS is provided by the observed radii of CM %Dra and YY Gem, as will be examined in \S 5.
%****
The SCVH EOS has been used previously to derive interior models for
solar (Chabrier et al. 1992; Guillot et al. 1995) and extrasolar (Saumon et al. 1996) giant planets and for brown dwarfs (Burrows et al. 1993).

%In the present section, we illustrate the consequence of the SCVH EOS on VLMS %interior
%models, namely $M=0.1$ $\msol$ and $M=0.6$ $\msol$. Comparison is made with %models based on MH EOS and FGVH EOS, for
%substantial difference is found already from the theoretical viewpoint with MM %EOS (see SCVH Fig. 21-23 or Saumon 1994).

The SC EOS is a pure hydrogen and helium EOS, based on the so-called additive-volume-law (AVL) between the pure components (H and He). The accuracy of the AVL has been examined in detail by Fontaine, Graboske \& VanHorn (1977).
The invalidity of the AVL to describe accurately the thermodynamic
properties of the mixture is significant only in the {\it partial ionization} region (see e.g. SCVH). As mentioned above, this concerns only a few percents of the stellar mass
under LMS conditions. Given the negligible {\it number}-abundance
of metals in stars ($\sim 2\%$ by mass, i.e. $\sim 0.2 \%$ by number)
 we
expect the presence of metals to be inconsequential on the
EOS. Their contribution to the perfect gas term is just proportional to the
number density ($P_{id}=\Sigma_i N_i kT/V$), i.e. $\sim 0.2\%$, whereas their
non-ideal (correlation) contribution
can be estimated either by the Debye-Huckel correction ($P_c\propto
(\Sigma_i N_i Z_i^2)^{3/2}$) for $\Gamma<1$ or by the electrostatic
(ion-sphere) term ($P_c\propto \Sigma_i N_iZ_i^{5/3}$) for $\Gamma>1$.
For solar metal-abundance (see e.g. Grevesse \& Noels, 1993) this yields an
estimated contribution to the EOS of $\sim 1\%$ compared with the
hydrogen+helium contribution.

Although this simple estimation shows 
that metals do not contribute appreciably
to the EOS of VLMS, as long as the {\it structure} and
the {\it evolution} are concerned,
we have decided to conduct complete calculations by comparing
models derived with the SC EOS and models based on the so-called MHD EOS
(Hummer \& Mihalas, 1988; Mihalas, Hummer \& D\"appen, 1988), which includes the contribution of heavy elements under solar abundances.
Although the MHD EOS is devoted to stellar {\it envelopes} and weakly
correlated plasmas, like the solar interior ($\Gamma \sim 0.1$), and thus
can not be applied to VLMS, it provides a useful tool for the
present test. The test is even strengthened by comparing the complete
MHD EOS with the pure hydrogen-helium ($Z=0$) MHD EOS kindly provided by
W. D\"appen. Note that the MHD EOS for mixtures does not assume the
additive-volume law between the various components, so that comparison with this
EOS provides also a test for the validity of this approximation.

Figure 1 shows in a HR diagram the results obtained with the MHD
EOS for solar metallicity (open circles) and for $Z=0$ (triangles)
 up to 1 $\msol$ \footnote{the 1 $\msol$ case is
shown only to illustrate the effect of metals in the EOS for the Sun; it is not
intented to reproduce an accurate solar model, since we use
a mixing length l$_{mix}$ = H$_p$ in the present calculations}.
The difference 
is less than 1\% in $\te$ and 4\% in L. This demonstrates convincingly the negligible contribution of the
metals to the
EOS over the entire LMS mass range.
As shown above, the contribution of metals to the EOS is proportional
to a power of the charge $Z$ and the atomic mass $A$.
Therefore, when applying a metal free EOS to solar metallicity
objects, the presence of metals can be mimicked by an equivalent helium fraction
$Y^\prime = Y+Z$ in the EOS, at fixed hydrogen abundance. Varying X instead
of Y would yield larger differences in $\te$ and $L$.
%results obtained with the zero-metal MHD EOS, but for the same Y as in the
%solar case (i.e X = 1 - 0.275) (cross on the figure).The differece between
%the zero-metal and
%the metal-rich case is by less than 2\% in $\te$, but can reach 16\% in L
Figure 1 also displays results based on the SC EOS, with the afore-mentioned equivalent
helium fraction (filled circles). Both the SC and MHD EOS yield very similar results.
The differences between the SC and the complete ($Z=Z_\odot$) 
MHD EOS amount to
less than 1.3\% in $\te$ and 1\% in $L$.
Below 0.3 $\msol$, however, the track
based on the MHD EOS starts to deviate substantially from the one based on
the SC EOS, a fairly reasonable limit for an EOS primarily devoted to solar
conditions.
\begin{figure}
%\picplace{2.5cm}
\epsfxsize=88mm
\epsfysize=80mm
\epsfbox{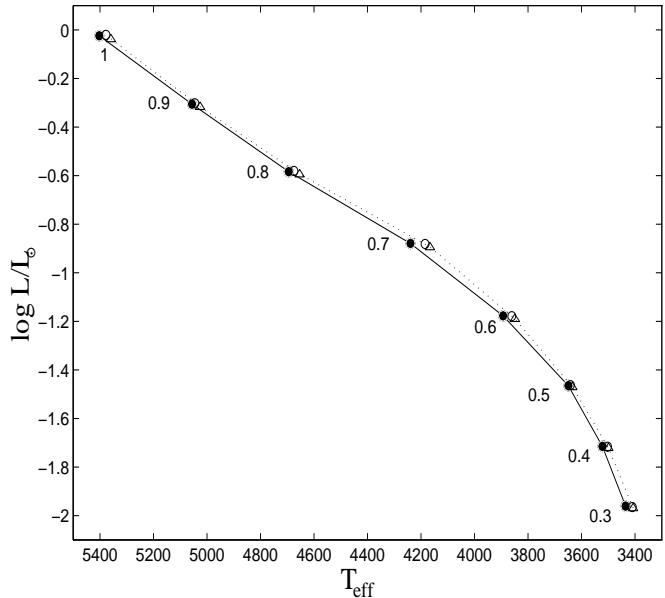} % where you want to insert a vbox for a figure
\caption[ ]{Theoretical HR-diagram for solar metallicity at
t=5 Gyrs.
Filled circles (solid line) Saumon-Chabrier EOS (SCVH, 1995)
with $Y\prime=Y+Z$; triangles (dotted line):
metal free (Z=0) MHD EOS with $Y\prime =Y+Z$;
open circles : solar metallicity ($Z=\Zsol$) MHD EOS (Mihalas et al. 1988).}
\end{figure}
This is better illustrated in Figure 2, which displays the adiabatic gradient
along the density-profile in a 0.2 $\msol$ and a 0.6 $\msol$ star
for the SC and MHD EOS. For the 0.6 $\msol$ star, slight
discrepancies between the SC and MHD adiabatic gradient appear in the
regime of partial ionisation of hydrogen 
and helium (log $\rho$ $\approx$ -4 to -0.5 and log T $\approx$ 4.2 to 5.3 ).
%and helium
%(log $\rho \approx$ -3.5 to -0.3
%and log T $\approx$ 4.3 to 5.8).
The differences become substantial for the 0.2 $\msol$ star because
of the strong departure from ideality in a large part of the interior.
The main discrepancies appear for log $\rho \wig > -3 ~g \, cm^{-3}$, with
log T $\approx 4$ and $\Gamma \approx ~2-3$, which marks the onset of hydrogen pressure and temperature partial ionisation. The unphysical negative value of the adiabatic gradient
in the MHD EOS for log $\rho > ~-1$ clearly illustrates the 
invalidity of the MHD EOS to describe the interior of dense systems, as
clearly stated by its authors (see Hummer \& Mihalas, 1988). Similar discrepancies occur also for other LMS EOS (see Saumon 1994; SCVH), which will
affect substantially the structure and the {\it evolution} of these convective objects.
We emphasize the excellent agreement between both EOS in the domain of validity of the MHD EOS,
even though the SC EOS does not include heavy elements. This clearly demonstrates the negligible
effect of metals on the adiabatic gradient. This latter is essentially determined by hydrogen and
helium pressure- and temperature- ionisation and/or molecular dissociation.
\begin{figure}
%\picplace{2.5cm}
\epsfxsize=88mm
\epsfysize=80mm
\epsfbox{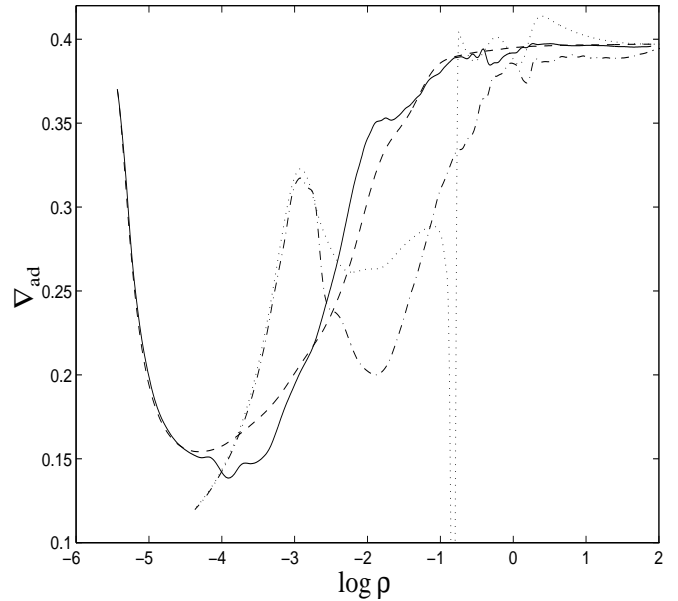} % where you want to insert a vbox for a figure
\caption[ ]{Adiabatic gradient as a function of density for the structure of
a 0.6 $\msol$ (solid and dashed curves) and 0.1 $\msol$ (dash-dot and dotted
curves).
The solid and dash-dotted curves correspond to the Saumon-Chabrier EOS (SCVH, 1995). The dashed and dotted curves
to the MHD (1988) EOS.
}
\end{figure}
These calculations clearly demonstrate that metal-free EOS can be safely
used to describe the structure and the evolution of LMS with a solar
abundance of heavy elements, providing the use of
an effective helium fraction to
mimic the effect of metals. It also assesses the validity of the Saumon-Chabrier EOS,
devoted primarily to dense and cool objects,
for solar-like masses\footnote{Only in term
of structure and evolution. For helioseismological studies, which require
extremely high accuracy ($\sim 1\%$ on the {\it speed of sound}), the effect of metals can become important and
the MHD EOS or the OPAL EOS (Rogers \& Iglesias, 1992),
specially devoted to such study, must be used.}.
% The MHD EOS
%is found to yield accurate results down to $\sim 0.4\,\msol$ but becomes
%invalid below this limit.

\subsection{The nuclear reaction rates}

The thermonuclear processes relevant from the energetic viewpoint under the central temperatures and densities characteristic
of VLMS are given by the PPI chain :

$$p+p\rightarrow d+e^++\nu_e   \eqno(1)$$
$$p+d\rightarrow ^3He\,+\,\gamma  \eqno(2)$$
$$^3He+^3He\rightarrow ^4He\,+\, 2p  \eqno(3)$$

%Central temperatures are too low to ignite helium-burning (Z=2).
The destruction of $^3He$ by reaction (3) is important only for
T $> 6\times 10^6$ K i.e masses M $\wig> 0.15\, \msol$ for ages $t<$ 10 Gyrs, 
since the lifetime 
of this isotope against destruction
becomes eventually smaller than a few Gyrs.
In the present calculations, we have also examined the burning
of light elements,
namely Li, Be and B, whose abundances
provide a powerfull diagnostic to identify
the mass of VLMS and brown dwarfs (see \S3.3).
Our nuclear network includes the main nuclear-burning reactions of
$^6Li,^7Li,^7Be,^9Be,^{10}B$ and $^{11}B$ (cf. Nelson et al. 1993).
We will focus on the depletion of
 the most abundant isotopes $^7Li,^9Be$ and $^{11}B$, described by the following
reactions :

$$^7Li\,+p\,\rightarrow \,2\,^4He  \eqno(4)$$
$$^9Be\,+\,p\,\rightarrow \, d\,+\,2\, ^4He \eqno(5)$$
$$^{11}B\,+\,p\,\rightarrow \,3\, ^4He \eqno(6)$$

The rates for these reactions are taken from
 Caughlan and Fowler (1988). These rates correspond to the reactions in the
vacuum, or in an almost perfect gas where kinetic energy largely dominates the
interaction energy. As already mentioned, such conditions are inappropriate for VLMS.
In dense
plasmas, the strongly correlated surrounding particles act collectively to
screen the bare Coulomb repulsion between two fusing particles. This will
favor the reaction and then enhance substantially the reaction rate with respect
to its value in the vaccuum. As mentioned in \S2.1, under the central conditions characteristic of low-mass
stars, the electrons are only partially degenerate and are polarized by
the ionic field. This responsive electronic background will also
screen the nuclear reactions and must be included in the calculations.
\footnote{This is what is called {\it electron screening}, the previous one
denoting the {\it ion screening}. We stress that there is some confusion in the literature, including in some textbooks, where the term electron screening is erroneously used to denote what is just the ion screening.}
Under the conditions of interest, both enhancement factors, ionic and
electronic, are of the same order, i.e. a few units (Chabrier 1997).
Different treatments of these enhancement factors have been derived, again following
the pioneering work of Salpeter (1954). A complete treatment of the {\it ionic}
screening contribution over the whole stellar interior density-range,
from the low-density Debye-Huckel limit to the high-density
ion-sphere limit was first derived by DeWitt et al. (1973) and
Graboske et al. (1973). The inclusion
of {\it electron} polarisability, in the limit of {\it strongly degenerate}
electrons, i.e. $\rs \rightarrow 0$ and $\psi \rightarrow 0$, was performed
by Yakovlev and Shalybkov (1989). An improved treatment of the ionic factor, and the
extension of the electron response to finite degeneracy, as found in the interior
of VLMS, was performed recently by Chabrier (1997).
The difference between the Graboske et al. and the Chabrier results, which illustrates both the improvement in the calculation of the ionic factor
and the effect
of electron polarisability, is shown on Figure 3 along temperature profiles
characteristic of VLMS, for two central densities, for Li-burning ($Z_1=1,
Z_2=3$). Substantial differences
appear, in particular in the intermediate-screening regime ($\Gamma \sim 1-10$)
characteristic of LMS and BD interiors.
%The values of the factors are given in Table 1.
The larger the charge, the larger the effect
($H\propto \Gamma \propto Z_1Z_2$). Such differences translate into differences in the abundances as a function of time and mass,
as will be examined in \S 3.3.
Note that the inclusion of electron polarisability was
found to decrease substantially the deuterium-burning minimum mass
(Saumon et al. 1996).
\begin{figure}
%\picplace{2.5cm}
\epsfxsize=88mm
\epsfysize=80mm
\epsfbox{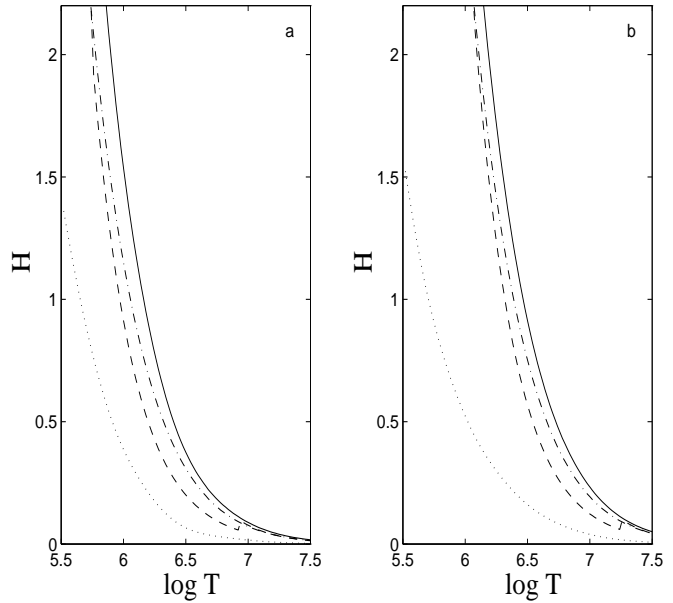} % where you want to insert a vbox for a figure
\caption[ ]{
 Screening factors as a function of temperature for
Lithium-burning (eqn. (4)), for
({\bf a}) $\rho=10$ g cm$^{-3}$ and
({\bf b}) $\rho=100$ g cm$^{-3}$. Solid line : present ion+electron screening
factors (Chabrier, 1997); dash-dot : ionic factor only; dot : electron factor
only; dashed line : Graboske et al. (1973).}
\end{figure}
\subsection{Deuterium-burning on the main sequence}

The {\it initial} D-burning phase ends after $\sim 10^6$ yrs
and is inconsequential for the rest of the evolution and the position on
the Main Sequence (Burrows, Hubbard \& Lunine, 1989 and \S 3 below).
We focus in this section on
the deuterium production/destruction rate along the PPI reactions, given
by eqn. (1) and (2), which is essential for the nuclear energy generation required to
reach thermal equilibrium. For stars below
$M \wig < 0.3 \, \msol$, which are entirely convective (see \S 3.2),
the deuterium lifetime against proton capture $\tau_{pd}$
is found to be much smaller than the mixing timescale in the central
regions, where energy production takes place.
The mixing time in these regions is estimated from the mixing-
length theory (MLT), $\tau_{mix}\approx l_{mix}/v_{mix}$, where
$v_{mix}\propto (\nabla - \nabla_{ad})^{1/2}$ is the mean
velocity of turbulent eddies.
These
fully convective stars are essentially adiabatic throughout most of their interior,
with a degree of superadiabaticity $(\nabla - \nabla_{ad})$ variing from $\sim$ 10$^{-8}$ to
$\sim 10^{-3}$ from the center to 99\% of the mass.
Thus the mixing-length parameter is inconsequential, and a
reasonable estimate for the mixing length is the pressure scale height
$l_{mix}\sim H_P$.
This yields a mixing timescale $\tau_{mix}\approx 10^7 - 10^8$ s, to be compared with
$\tau_{pd} << 10^6$ s
in the central part of the star where nuclear energy production takes place.
This is illustrated in Fig.4 where both timescales are compared in a 0.075 $\msol$ star evolving on the main sequence.
\begin{figure}
%\picplace{2.5cm}
\epsfxsize=88mm
\epsfysize=80mm
\epsfbox{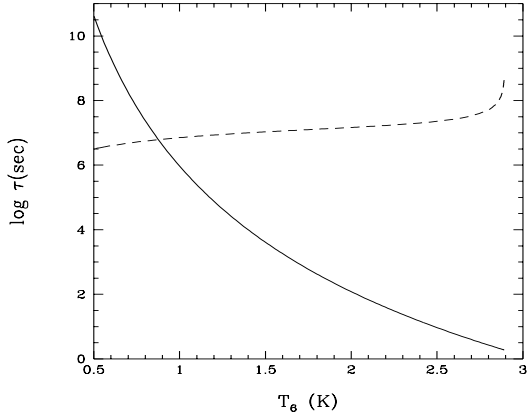} % where you want to insert a vbox for a figure
\caption[ ]{Comparison of the deuterium burning lifetime against
$p$-capture (solid line) and the mixing timescale (dashed line)
in a 0.075 $\mso$, with solar metallicity. Deuterium burning occurs in the region
$T> 10^6$ K.}
\end{figure}
Since deuterium is burned much more quickly than it is mixed, a deuterium abundance gradient will develop in the central layers.
This process can be described by the stationary solution
of the following diffusion equation,
since the diffusion and the nuclear timescales are orders of magnitudes smaller than the evolution time :

$$D_{mix} \nabla^2Y_d\,+\, \lambda_{pp} {Y_p^2 \over 2} \,-\, \lambda_{pd}
Y_pY_d  = 0 \eqno(7)$$

Here $D_{mix}\approx v_{mix}l_{mix} \propto 1/\tau_{mix}$ is the convective mixing diffusion coefficient,
$\lambda_{pp}$ and $\lambda_{pd}$ are the
respective rates of reactions (1) and (2), and $Y_p$ and $Y_d$ denote respectively the hydrogen and the deuterium abundances by number
($\tau_{pd}={1 \over Y_p\lambda_{pd}}$, following
the notations of Clayton 1968).
As long as $\tau_{pd} << \tau_{mix}$, the deuterium abundance $ Y_{d}$
in each burning layer will be
close to its nuclear quasi-equilibrium value, as given by $D_{mix}\sim 0$ in Eq.(7). The abundance of deuterium is relevant only in the central region, where
we adopt the equilibrium value. The complicated task of solving Eq.(7) is thus
not necessary in this case.
The nuclear energy production-rate of the $d(p,\gamma)$ reaction is then given by:

$$\epsilon_{pd}=  Q_{pd} \lambda_{pp} {Y_p^2 \over 2}\eqno(8)$$

where $Q_{pd}$ is the energy of the $d(p,\gamma)$ reaction.
Note that during the {\it initial} deuterium
burning phase, this situation does not occur and the deuterium abundance is calculated
like other species in the network,
under the usual {\it instantaneous mixing} approximation, which corresponds
to $\tau_{mix} << \tau_{nuc}$. In that case abundances are homogeneous
throughout the whole burning core, i.e. $\nabla Y_d=0$ and the concentration
of deuterium is given by an average value over the convective
zone $<Y_d>$. When $\tau_{nuc}< \tau_{mix}$, instantaneous mixing
will thus overestimate the deuterium-concentration in the central layers.
This can be understood intuitively
since instantaneous mixing will provide too much deuterium, i.e. more
than produced by nuclear equilibrium,
at the bottom, i.e. the hottest part, of the burning region.
This yields an overestimation of the nuclear energy production at a given temperature and $Y_p$, and thus
of the total luminosity of the star. This effect is not drastic for stars
M $ > 0.1 \msol$, but increases the luminosity by $\sim 20\%$ for 0.1 $\msol$
and by $\sim 55\%$ for 0.075 $\msol$ and thus bears important consequences
for a correct determination of the stellar to sub-stellar transition.

\subsection{Model atmospheres}

The low temperature and high pressure in the photosphere of M-dwarfs raise
severe
problems for the computation of accurate atmosphere models.
For these low effective temperatures ($\wig < 5000$ K)
molecules become stable (H$_2$, H$_2$O, TiO, VO,...), and constitute
the main source of absorption along characteristic wavelengths.
The presence of these molecular bands complicates tremendously the treatment of radiative transfer, not only because
of the numerous transitions to be included in the calculations, but also because the molecular
absorption coefficients strongly depend on the frequency and a grey-approximation, as used for more massive stars, is
no longer valid. Moreover, the high density in M-dwarf
atmospheres yields the presence of {\it collision-induced} absorption,
an extra degree of complication.
These points have been recognized long ago
and motivated various developments in the modelling of M-dwarf atmospheres since
the pioneering
work of Tsuji (1966). Substantial improvement in this field
has blossomed in recent years with the work of Allard and
collaborators (Allard 1990; Allard and Hauschildt 1995a; 1997), Brett (1995), Tsuji and collaborators (Tsuji et al. 1996)
and Saumon (Saumon et al. 1994), due to recent interest in the extreme lower main sequence and the
necessity to derive accurate atmophere models to identify the luminosity and the colors of M-dwarfs and brown dwarfs.
%Until recently, all evolutionary models relied on {\it grey} atmosphere %models, essentially based
%on the Rossland opacities of Alexander and collaborators.
%The first evolutionary models with non-grey atmosphere models
%were conducted by the Tucson group (Saumon et al. 1994), but for %zero-metallicity. Besides the fact
%that these models might apply to still hypothetical primordial Pop III %objects, they provide a useful
%limit (benchmark ?) to calculations at finite metallicity. The first %evolutionary models with
%non-grey atmosphere models at finite-Z are due to Baraffe et al. (1995), %and were shown to represent
%a substantial improvement with respect to previous models when compared %with observations of faint and red objects.
We refer the reader to the recent review by Allard et al. (1997b) for details.

The present evolutionary calculations are based on
the latest generation
of LMS non-grey atmosphere models at finite metallicity (Allard \& Hauschildt
1997; AH97), labeled $NextGen$.
In order to illustrate the most recent improvements in LMS atmosphere
theory, we will make comparisons with stellar models based on the previous
so-called {\it Base} models (Allard and Hauschildt 1995a; AH95), as used in
the calculations of Baraffe et al. (1995). 
A preliminary version of the {\it NextGen} models was used by Chabrier, Baraffe \& Plez (1996) and Baraffe \& Chabrier (1996) and compared with results based
on the other source
of non-grey atmosphere models presently available, computed by Brett and Plez
(Brett 1995; Plez 1995, private communication; BP95). This latter set, however, is restricted
to solar metallicity.
As shown in Chabrier et al. (1996) the  BP95 models lead to
$\te$ intermediate between the $Base$ and $NextGen$ models, for a given mass.
Note that the AH97 models 
used in the present work
include 
improved molecular opacity treatment compared to the straight mean method used in the {\it Base} models, and
the more recent water linelist
of Miller et al.  (1994) (see AH97 for details). A summary of the main differences in the input of these atmosphere models is outlined in Chabrier et al. (1996).

%It is important to stress that the AH and the BP models stem from {\it completely}
%independent works, not only for the input opacities but also for the
%resolution of the transfer equations.
%The main differences between these different sets of models stem essentially from : i) different
%technics in the opacity calculations : the AH95 models were based on the Straight-Mean (SM) approximation,
%whereas BP95  and AH97 use the more accurate Opacity-Sampling (OS) technique; ii) different linelists,
%in particular for TiO, the main source of absorption below $\te \sim 3500$ K in the optical :
%the SM coefficients by Collins (1975) in {\it Base}, the linelist from Jorgensen (1994) in
%{\it NG2} and $NextGen$ and the linelist computed by Plez, Brett \& Nordlund (1992) for BP95;
%the oscillator
%strengths are fitted to observed spectra in AH95, they are based on laboratory experimental data in BP95
%whereas AH97 ??;
%iii) BP95 use a pseudo linelist for H$_2$O based on old SM data (cf Plez, Brett and Nordlund
%1992), while AH95 use the original SM tables (see AH95), the more recent OS tables
%from Jorgensen (1994)  for {\it NG2} and the Miller et al. (1994) linelist
%for  $\NeG$.
As shown e.g. by Jones et al. (1995),
the models still predict too strong infrared water bands despite the inclusion of these new (even if still incomplete and
preliminary) water linelists (see AH97).
This shortcoming, and the remaining uncertainties in the calculation of the
$TiO$ absorption coefficients, represent the main limitation of
present VLMS atmosphere models for solar-like metallicities. A second limitation comes from grain formation below
$\te < 2200 K$ (Tsuji, Ohnaka \& Aoki; 1996) which is likely to affect
the spectra and the atmosphere structure of the coolest
M-dwarfs, and brown dwarfs,
for solar-metallicity. Work in this direction is under progress.
%Models of AH are available for a wide range of metallicities whereas the %BP models have been calculated
%on an extended range of effective temperatures only for solar metallicity

\subsection{Boundary conditions}

The last but not least problem arising in the modelization of VLMS is the determination of accurate
outer boundary conditions (BC) to solve the set of internal structure equations.
All previous VLMS models relied on grey atmosphere models.
The BC were based either on a $T(\tau$) relationship
(Burrows et al. 1989; Dorman et al. 1989; D'Antona \& Mazzitelli 1994, 1996;
Alexander et al. 1996) or
were obtained by solving the radiative transfer equations
(Burrows et al. 1993). In order to make consistent comparison with these
models, and to demonstrate the limits of a grey approximation for VLMS, we first give a short overview of the various procedures
used in the literature.

$T(\tau$) relationships have the generic form :
$$T^4= {3\over4} \te^4 (\tau + q(\tau))\eqno(9)$$
with different possible q($\tau$) functions (Mihalas, 1978).
The most simple form is based on the Eddington approximation,
which assumes that the radiation
field is isotropic, which yields q($\tau$)=2/3, as used in Burrows et al.
(1989).
An exact solution of the grey  problem (Mihalas 1978) gives actually a function which departs slightly from 2/3, but this correction is
inconsequential on the resulting evolutionary models (Baraffe \& Chabrier 1995).

%approximation with the boundary condition obtained analytically by a %single-layer
%integration from $\tau$ = 0 to $\tau$ = 2/3, defined as the surface of the %star (Appenzeller 1972).
%This approach was used by Baraffe et al.(1995) to compare their non-grey
%models to grey models. Though extremely rough, this procedure gave
%amazingly results which did not depart drastically from non-grey models
%{\it for solar metallicity}. However, for lower
%metallicties, this prescription is definitly wrong and predicts an %erroneous mass - $\te$ relationship..
%The usual procedure consists of constructing atmosphere models and using %the conditions (T,P or T,$\rho$) at a given value of the optical depth
%($\tau_{bc} > 1$) as boundary conditions for the inner integration.
%The first (and most difficult) task is to elaborate the model atmosphere %and the second is to choose $\tau_{bc}$.
As mentioned in the previous sub-section, the strong frequency-dependence
of the molecular absorption coefficients yields synthetic spectra
which depart severely from a frequency-averaged energy distribution
(see e.g. Allard 1990; Saumon et al. 1994).
Modifications
of the function q($\tau$) have been derived in the past in order to mimic
departures from greyness
(Henyey et al. 1965, as used by D'Antona \& Mazzitelli 1994; Krishna-Swamy 1966, as used by Dorman et al. 1989). These corrections, however, are
based at least partly on ad-hoc calibrations to the Sun and do not rely on reliable grounds.

The assumption of a temperature stratification following a $T(\tau$)
relationship requires not only a grey approximation {\it but also} the assumption of radiative equilibrium, implying that all the energy in the optically thin layers is transported by radiation.
However, below $\sim 5000$ K, molecular hydrogen recombination in the envelope (H+H $\rightarrow$ H$_2$) reduces the entropy and thus the adiabatic gradient (see SCVH). This favors the onset of
convective instability in the atmosphere so that
convection penetrates deeply into the optically thin layers
(Auman 1969; Dorman et al. 1989; Allard 1990; Saumon et al. 1994; Baraffe et al. 1995).
Radiative equilibrium is no longer satisfied
and flux conservation in the atmosphere now reads
$\nabla(F_{rad}+F_{conv})=0$.
Though rigorously inconsistent with the use of a $T(\tau)$ relationship,
by definition, modifications of eqn.(9) have been proposed to
account for convective transport in the optically thin layers.
Henyey et al. (1965) prescribed a correction to the calculation of the
temperature gradient and to the convective efficiency
in optically thin layers which is equivalent to a correction of
the diffusion approximation. This procedure was used by Dorman et al. (1989). On the other hand, some authors have just neglected the
presence of convection in
the optically thin regions of the atmosphere (Alexander et al. 1996).
%As will be shown below, taking convection into account with
%the use of a T($\tau$) relationship will overestimate  the effective
% temperature at a given mass, the effect increasing with  decreasing %metallicities.

All these arguments show convincingly that a description of M-dwarf atmospheres
based on grey models and $T(\tau)$ relationships is physically  incorrect.
The consequence on the {\it evolution} and the {\it mass-calibration}
can be determined by comparing stellar models
based on the various afore-mentioned grey treatments with the ones
based on non-grey atmosphere models and proper BCs.
These latter are calculated as follows.
We first generate 2D-splines
of the atmosphere temperature-density profiles in a ($\log g, \te$)-plane, for a given metallicity. The connection between
the atmosphere and the interior profiles is made at $\tau = 100$,
the corresponding (T-$\rho$) values being used as the BC for the Henyey integration. This choice is motivated by the fact that
i) at this optical depth, {\it all} atmophere models are adiabatic and can be matched with the interior adiabat,
and ii) $\tau= 100$ corresponds to a photospheric radius $R_{ph}\,<\, 0.01\, R_\star$, where $R_\star$ is the stellar
radius, so that the Stefan-Boltzman equation $\te^4=L_\star/4\pi\sigma R_\star^2$ holds accurately. We verified that varying this BC
from $\tau\approx 30$ to 100 does not affect significantly the results.
Convection starts to dominate,
i.e carries more than 50\% of the energy in the atmosphere for
$\tau \wig > 1$ (Allard, 1990; Brett, 1995). Therefore, $\tau$ = 100 is a safe
limit to avoid discrepancy between the treatment of convection
in the atmosphere and in the interior.
Note that $\tau=100$ corresponds to a pressure range
$P\approx 10^{-1}$ to $  10^3 $ bar, depending on the temperature, gravity and
metallicity. Along this important pressure-range,
the dominant source of absorption shifts, going inward in the atmosphere, from water absorption to TiO,
and eventually CIA H$_2$ absorption,
whereas the molecular line width changes from thermal-broadening to
pressure broadening (AH95; Brett, 1995). For a fixed mass and composition,
there is only one
atmosphere temperature-density (or pressure) profile with a given effective temperature and gravity
wich matches the interior profile
for the afore-mentioned BC. This determines the complete stellar model for
this mass and composition.
The effective temperature, the colors and the bolometric corrections are given by the atmosphere model.
%(cf. Burrows et al. 1989) and on the T - $\tau$ relationship of %Krishna-Swamy (1966), as this latter is mostly used. In this case,
%the correction to the diffusion approximation for the calculation of the %temperature
%gradient (Henyey et al. 1965) is used and convection in the optically thin %region is taken into account, in order to compare our results with those
%of Dorman et al. (1989). We will however also consider the case where %convection is neglected in regions where T $\le \te$, as the departure %from results obtained with non-grey models is amazingly small.

Figures 5a-b show the temperature-pressure profile of $NextGen$ atmospheres for $\te$ = 3500 K and log g = 5 for [M/H]=0 (Fig. 5a) and [M/H]=-1 (Fig. 5b).
The location of the onset of convection is shown on the figures. For both metallicities, convection reaches the optically thin region
($\tau$ = 0.02 for [M/H]=0 and $\tau$ = 0.06 for [M/H]=-1). Also shown
are profiles derived from grey models and $T(\tau)$ relationships using
the Eddington approximation (dotted line), the
Krishna-Swamy (KS) relation with a correction for the presence of convection in optically thin layers (cf. Henyey et al. 1965) (dash-dot line) and with convection arbitrarily
 stopped at
T$\le \te$ (dashed line).
The Eddington approximation yields severely erroneous results. The atmosphere profile is substantially cooler
and denser than the non-grey
one above $\tau \sim 1$, so that a hotter atmosphere model is required to
match the internal adiabat, yielding too hot stellar models, as demonstrated
in Chabrier et al. (1996). For solar metallicity, the two profiles based on the KS relation are almost undistinguishable and yield atmosphere models very close to the non-grey one.
%This stems from the fact that, even though
%convection starts at $\tau << 1$, it is inefficient in the optically thin layers for solar metallicity, and does not
%modify appreciably the temperature gradient.
This agreement, however, strongly depends on the
thermodynamic conditions, pressure and temperature, along the
atmosphere profile. This is shown convincingly in figure 5b, for a denser metal-poor atmosphere.
The higher pressure of metal-poor models favors the convective flux
in the present grey models,
and thus yields a flatter temperature gradient.

%In the optically thin regions with convective instability, the density at
%[M/H]=-1 (resp. [M/H]=-2) is $\sim $ 5 times (resp. $\sim$ 20 times)
%larger than for solar metallicity at $\te$ = 3500 K and log g = 5.
In that case, both KS profiles depart substantially from the correct one, even though
the model based on the KS relation {\it without} 
convection in optically thin region, although physically inconsistent, yields less severe disagreement ($\sim 100$ K in $\te$).
It must be kept in mind, however, that the convection correction in a grey atmosphere based on a $T(\tau$) relationship does not
reflect adequately the influence of convection on a correct {\it non-grey} model
(see e.g. Brett 1995, AH95; AH97).
It also demonstrates that the Henyey et al. (1965) correction to
the radiative diffusion approximation, when using a $T(\tau)$ relationship, 
overestimates the convective flux in optically thin regions. This yieds
flatter temperature gradient in the atmosphere and thus larger effective temperature
for a given mass.
The models of Burrows et al. (1993), although based on grey
atmosphere models, do not rely on a $T(\tau)$ relation but use BC based on
the resolution of the transfer equations. Although cooler than the
Burrows et al. (1989) models, they still yield too large effective
temperatures compared with the non-grey models. As shown above this
stems very likely from an overestimation of convection efficiency in the
atmosphere.

Even when the temperature in the atmosphere is low enough so that
convection does not penetrate anymore into the optically thin region
($\te \wig < 2500 $K),
strong departure from greyness still invalidates the use of a $T(\tau)$ relationship. This is illustrated in Figure 5c for solar metallicity models with $\te$ = 2000 K and log g = 5.5.
\begin{figure}
%\picplace{2.5cm}
\epsfxsize=88mm
\epsfysize=50mm
\epsfbox{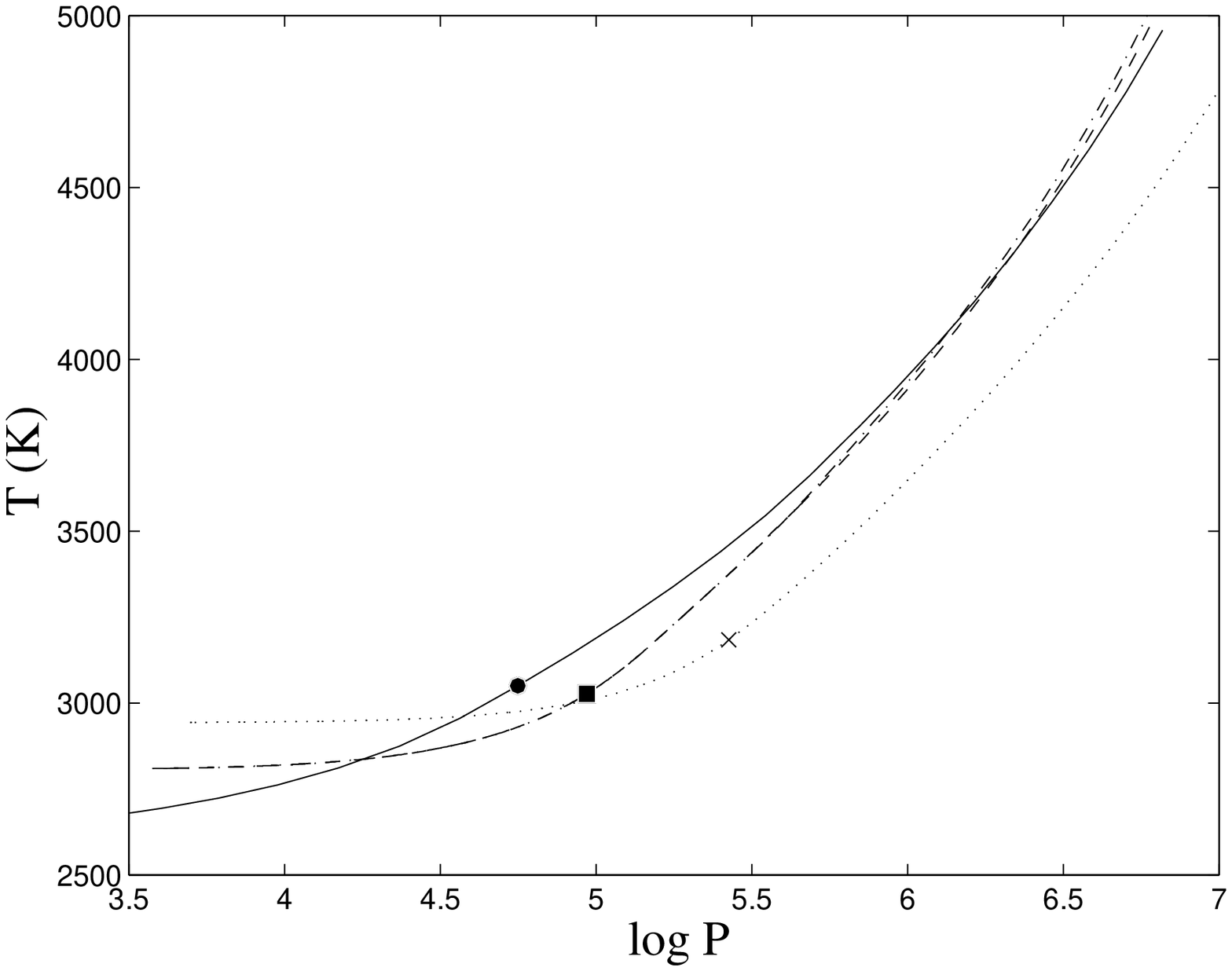} 
\epsfxsize=88mm
\epsfysize=50mm
\epsfbox{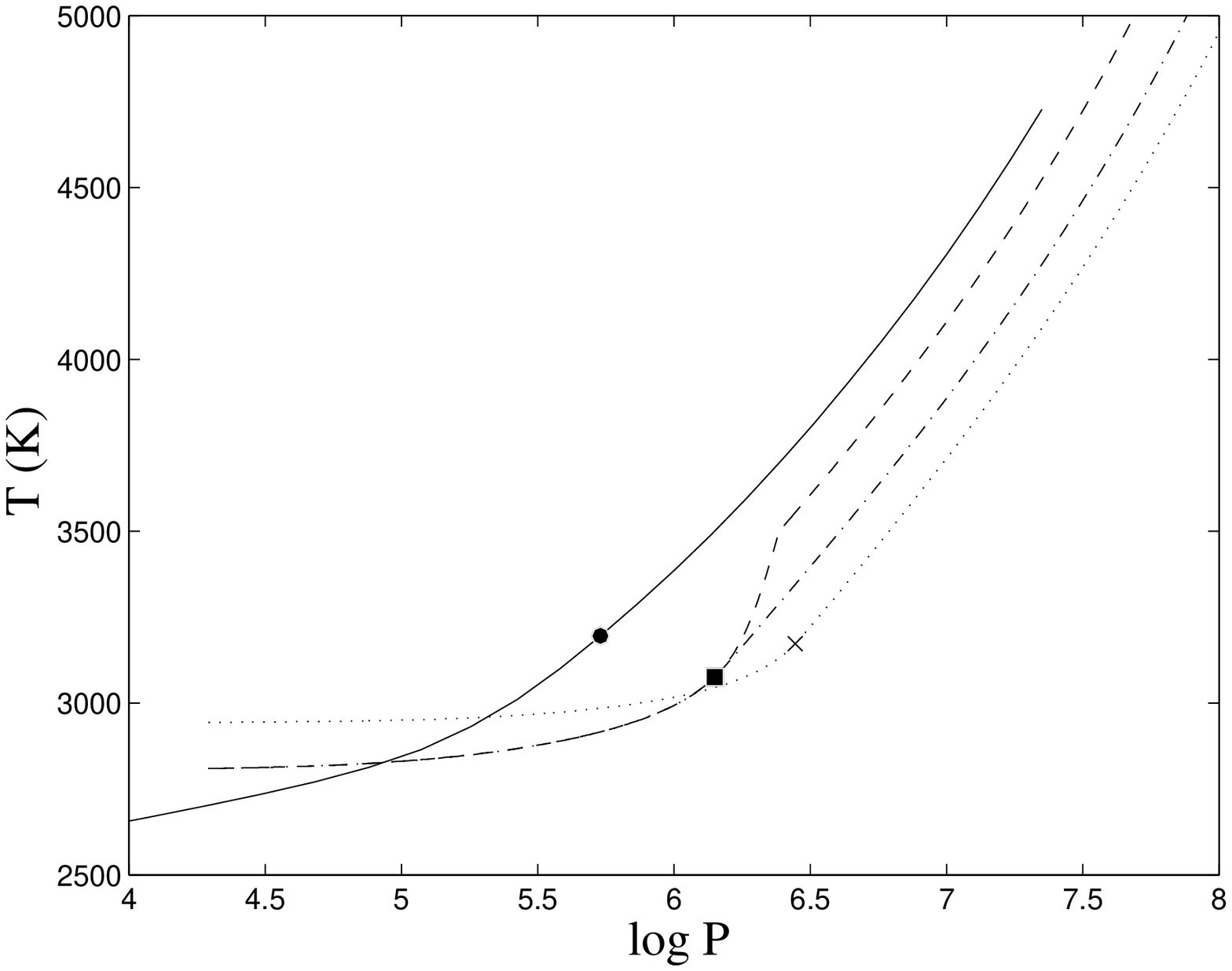}
\epsfxsize=88mm
\epsfysize=50mm
\epsfbox{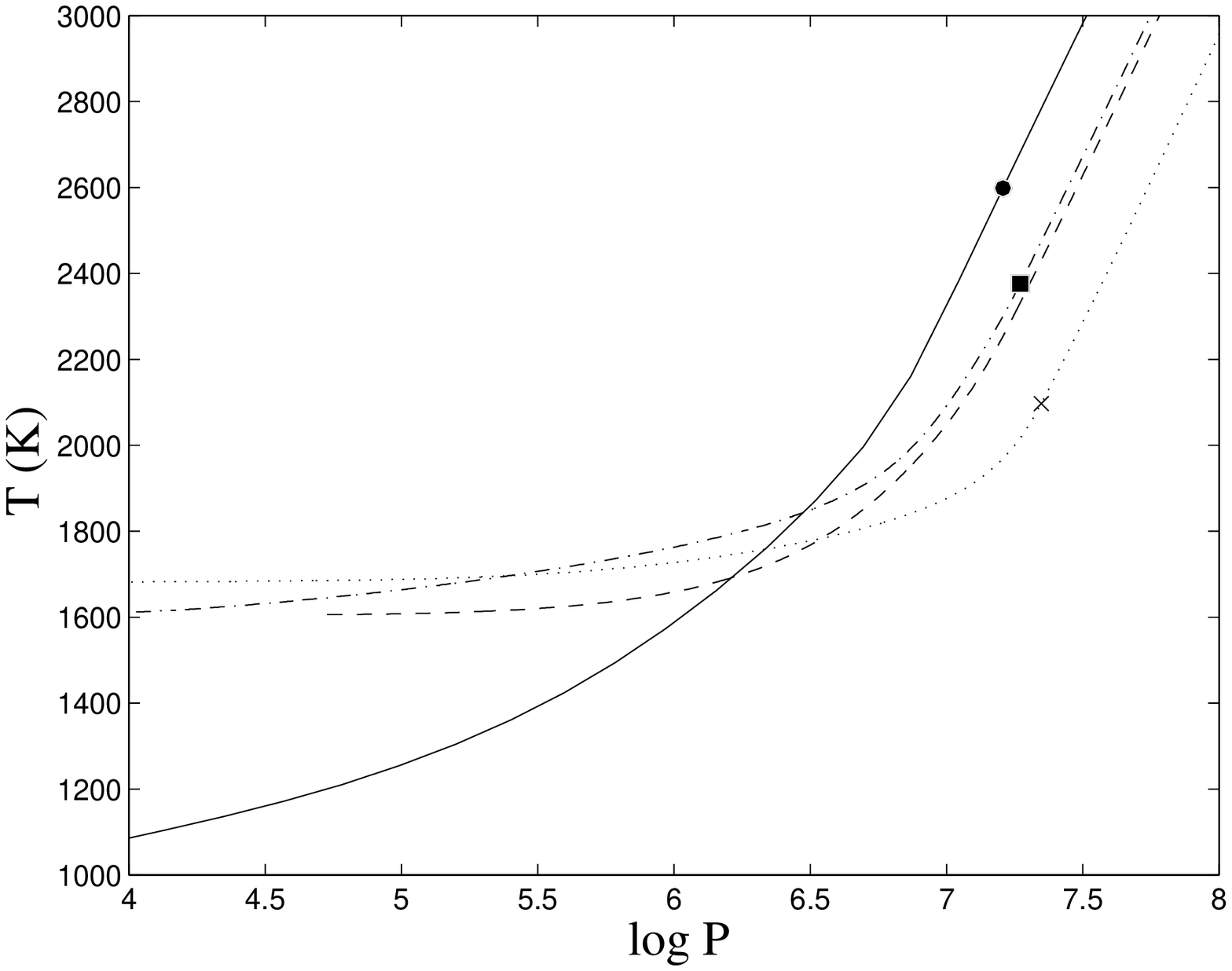}
\caption[ ]{$P-T$ atmosphere profile according to
the non-grey models of Allard \& Hauschildt (1997) (solid line). Dash-dotted line:
grey models obtained with the Krishna-Swamy (1966) T-$\tau$
relationship with convection included in the optically thin region.
Dashed line :  {\it idem} with convection arbitrarily stopped at T $\le \te$,
i.e. $\tau \sim 1$.
Dotted curve  : Eddington approximation. {\bf (a)}
$\te$ = 3500 K, $\log g = 5$, [M/H]=0.
The points indicate the onset of
convection, corresponding to $\tau$ = 0.02 for the non-grey model (circle),
$\tau$ = 0.06 for the Krishna-Swamy case (square) and $\tau$ = 0.25 for Eddington (cross). The dashed and dot-dashed curves are undistinguishable in that case.
For the dashed curve, convection sets in at T = $\te$, by definition.
{\bf (b)} $\te$ = 3500 K, log g =5,  [M/H]=-1. Convection occurs
 at $\tau$ = 0.06 for the non-grey model,
$\tau$ = 0.08 for the Krishna-Swamy case and $\tau$ = 0.24 for Eddington. Note
the departure between the dashed and dash-dotted curve when convection sets in in the optically thin region.
%, showing the high efficiency of atmospheric convection at lower metallicity.
{\bf (c)} $\te$ = 2000 K, log g =5.5, [M/H]=0.
In this case, the dashed curve corresponds to a model based on the Krishna-Swamy relation with {\it
grainless} Rosseland opacities (D. Alexander, private communication).
Convection starts well below the photosphere ($\tau > 1$) and the departure between grey and non-grey models stems essentially from strong non-grey effects.
}
\end{figure}

The effect of grain formation in the atmosphere on the evolution
was considered in Chabrier et al. (1996), by comparing, within a grey approximation,
stellar models based on
the Alexander and Fergusson (1994)
Rosseland opacities and on similar dust-free opacities kindly provided by Dave Alexander.
Grains were found to affect the evolution
only below $\sim 1800 K$. This is confirmed on
figure 5c where the two grey profiles, with and without grain, yield essentially the same atmospheric
structure at $\te=$2000, i.e. $\sim 0.075\,\msol$.
These calculations will be reconsidered once non-grey atmosphere models with grains will be available.

These calculations show convincingly that procedures based on a grey
approximation and a $T(\tau)$ relation for the derivation of VLMS evolutionary models
are extremely unreliable, even though they may
yield, under specific thermodynamic conditions, to (fortuitous) agreement
with consistent non-grey calculations. As a general result, a grey
treatment yields cooler and denser atmosphere profiles
below the photosphere (cf. Saumon et al. 1994; Allard and Hauschildt 1995b), and thus overestimates
the effective temperature for a given mass (Chabrier et al., 1996). Therefore, they lead to
erroneous mass-luminosity and mass-$\te$ relationships, as will be discussed in \S 4.

\section{Evolutionary tracks}

As mentioned in the introduction, the present study focusses on a mass-range limited to $M \le 0.8 \msol$ and will consider
the brown dwarf domain only scarcely. For larger masses, the physics
of VLMS remains unchanged but variations of the mixing length
parameter start to be consequential and require comparison with
observations (such comparisons are considered in detail in Baraffe et al. 1997). On the other hand,
a substantial improvement over existing brown dwarf models
(e.g. Burrows et al. 1989; 1993), requires the derivation of non-grey
atmosphere models {\it with grains}. As shown by the recent analysis of Tsuji et al. (1996),
silicate and iron grains can contribute significantly to the opacity in the photosphere below
$\sim 2200$ K (see also Lunine et al. 1986; Alexander and Fergusson 1994),
 a typical effective temperature for massive brown
dwarfs with solar abundances (see \S 3.1).
%Interestingly enough, the spectrum of the cool BD $Gl229B$ was better reproduce with
%{\it grainless} opacities, suggesting a sedimentation process below a certain temperature
%(Allard et al., 1996; Tsuji et al.; 1996). 
%As grains affects not only the spectral distribution i.e the
%resulting colors and magnitudes, but also
%the structure of the atmosphere, one may expect either an effect on the %evolution.

In the following sub-sections, we present the evolution of the
mechanical and thermal properties of objects ranging from 0.055 to 0.6 $\msol$, over a metallicity-range [M/H]= -2 to 0.
The general properties of VLMS and BDs have already
 been described by numerous authors (see \S 1).
 We will not redo such an analysis in detail
but rather focus on selected masses to illustrate the differences arising
from the new physics (EOS, nuclear rates, non-grey model atmospheres)
described in the previous section. The different mechanical and thermal properties of the
present models, for various metallicities, are presented
in Tables 2-7.

\subsection{Mechanical and thermal properties}

\subsubsection {Internal structure}

Figure 6 displays the behaviour of the central temperature, the radius
and the degeneracy parameter along the main sequence (MS) for two
metallicities, [M/H]=0 (solid line) and [M/H]=-1.5 (dashed line).
For {\it stars} on the MS, the internal temperature is
large enough in the stellar interior for the pressure to
be dominated by {\it classical} contributions ($P=\rho kT/\mu m_H$),
so that hydrostatic equilibrium yields $R\propto M/T$.
%In fact, below $\0.4\,\msol$, the star becomes entirely convective
%(see \S 3.2), which yields the well known polytropic relation $R\propto M??$.
Below $\sim 0.15 \, \msol$,
the object is dense and cool
enough for the electrons to become substantially degenerate ($\psi \propto T/\rho^{2/3}$), so that the electronic {\it quantum}
contribution ($P\propto \rho^{5/3}$) overwhelms the ionic classical pressure. Eventually this will yield the well-known mass-radius
relation for fully degenerate (zero-temperature) objects $R\propto M^{-1/3}$.
%Detailed calculations show that the full degeneracy limit occurs for
%$M\approx 0.065\,\msol\,=\,6.5\,M_J$, where $M_J$ denotes the Jupiter mass (Saumon et al. 1996).
The brown dwarf domain lies between these two limits
(classical and fully degenerate) and is characterized by $R\sim R_0(1+\psi)$
$\sim 10^{-1}\,R_\odot$, about Jupiter radius,
where $R_0$ is the zero-temperature (fully degenerate) radius (see e.g. Stevenson, 1991).
 The transition between
the stellar and sub-stellar domains
is characterized by this ongoing electron degeneracy in the interior, as illustrated
in Figure 6.
Note also that once degeneracy sets in, the temperature scales as
$T\sim M/R-M^{2/3}/R^2$, where the first and second term represent the classical
and quantum gas contributions, respectively. This yields the rapid drop of
the interior temperature (and effective temperature for a given $\te-T_{int}$
relation) near the sub-stellar transition, and thus
the characteristic severe drop in the luminosity
($L\propto R^2\te^4$).
\begin{figure}
%\picplace{2.5cm}
\epsfxsize=88mm
\epsfysize=80mm
\epsfbox{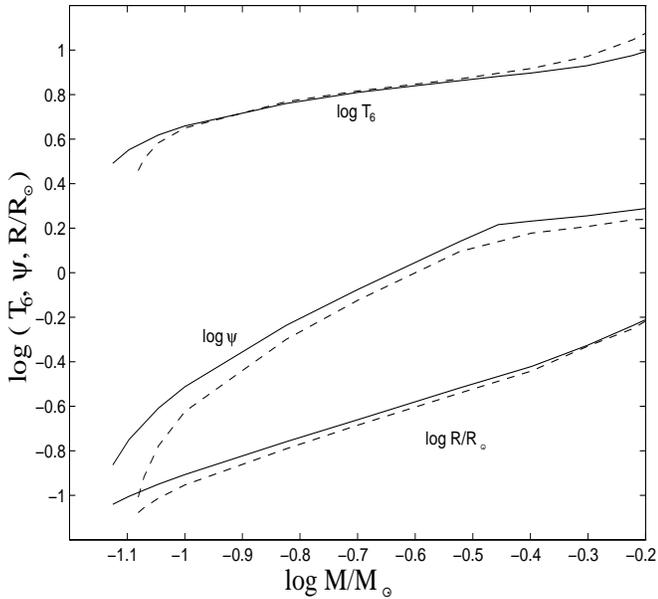} % where you want to insert a vbox for a figure
\caption[ ]{Degeneracy parameter $\psi=T/T_F$, central temperature $T_{C}$ (in units of $10^6$ K) and
radius ($R/R_\odot$),
as function of mass for
[M/H]=0 (solid line) and [M/H]=-1.5 (dash).}
\end{figure}

Figure 7 displays the evolution of the radius for different masses.
% with [M/H]=0 (Fig. 11a) and [M/H]=-1.5 (Fig. 11b).
Although the radius is fixed mainly by the EOS, it does depend, to some extent, on the atmosphere treatment. The
effect is negligible for solar metallicity,  as shown on the figure for $m=0.2 \,\mso$, but can yield $\sim 3\%$ difference on
the final radius for $\mh = -1.5$. 
%MAXIMUM EFFECT ***
After a similar pre-MS contraction phase for all masses,
the hydrogen-burning stars reach hydrostatic equilibrium whereas objects
below the hydrogen-burning minimum mass keep contracting until
reaching eventually the afore-mentioned asymptotic radius, characteristic of
a strongly degenerate interior.
%The present models have been shown to reproduce accurately the observed radii of
%the low-mass eclipsing binary systems CM-Dra and YY-Gem (Chabrier \& Baraffe, %1995).\footnote{The radii based on the present atmosphere models are $\sim %1.5\%$ smaller than the ones obtained in Chabrier \& Baraffe, 1995, based on
%the {\it Base} atmosphere models.}
\begin{figure}
%\picplace{2.5cm}
\epsfxsize=88mm
\epsfysize=80mm
\epsfbox{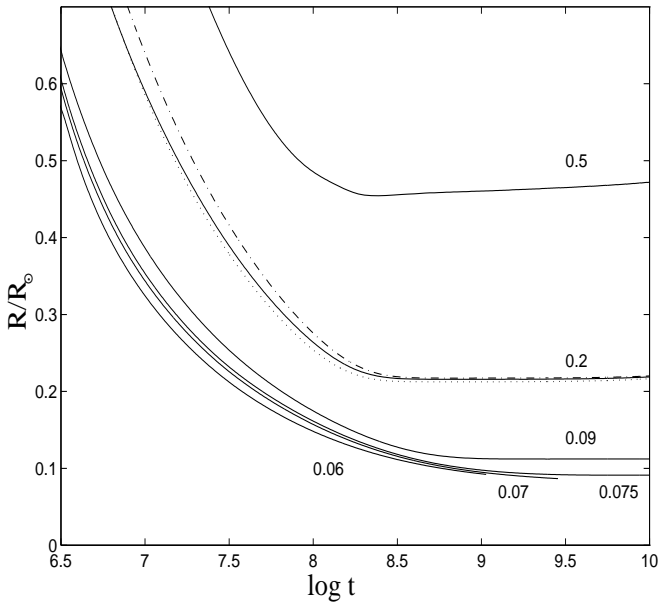} % where you want to insert a vbox for a figure
\caption[ ]{Evolution of the radius (in solar unit)
for different masses, for solar metallicity.
For 0.2 $\msol$, comparison is shown with results obtained with the
Krishna-Swamy T-$\tau$ relationship (dash-dotted curve) and with the Eddington approximation (dotted line).}
\end{figure}

\subsubsection {Luminosity}

Figures 8a-b 
exhibit $L(t)$ for different
masses, for $\mh=0$ and $\mh=-1.5$, respectively. Initial deuterium burning proceeds very quickly,
at the very early stages of evolution, and lasts about $\sim 10^6-10^7$ years. Our
calculations were done with an initial deuterium abundance $[d]_0=2 \times 10^{-5}$ in mass fraction, which corresponds to the average
abundance in the interstellar matter (cf. Linsky et al. 1993). A value $[d]_0=5 \times 10^{-5}$ increases
the deuterium burning timescale by a factor $\sim$ 2 and the luminosity by
10\% to 50\% during this phase. We verified that this effect is inconsequential for the rest of the evolution.
As clearly shown on the figures, for masses above $0.07\,\msol$ for [M/H]=0 and
$0.08\,\msol$ for [M/H]=-1.5, the internal energy
provided by nuclear burning quickly balances the contraction gravitational energy, and the lowest-mass star reaches
complete thermal equilibrium ($L = \int \epsilon dm$, where $\epsilon$ is the nuclear energy rate), after $\sim 1$ Gyr, for both
metallicities.
The lowest mass
for which thermal equilibrium is reached defines the so-called hydrogen-burning minimum mass (HBMM), and the related
hydrogen-burning minimum luminosity (HBML). These values are given
in Tables 2-7, for various metallicities.
%which is $log\, L/\lsol \sim -4.0$ for $\mh=0$ (M= 0.075 $\msol$) and
%$log L/\lsol=-3.8$ for $\mh=-1.5$ (M = 0.083 $\msol$).
%In the range [M/H]=-1.5  to -1, we find that the 0.08 $\mso$ become %browndwarfs and  0.083 $\mso$
%reach thermal equilibrium. Whereas for [M/H]=-2, the 0.083 $\mso$  is just %a the limit of the HBBM and slowly cools down.
Note the quick decrease of luminosity with time for objects below the HBMM, with
$L\propto 1/t$ (cf. Burrows et al. 1989; Stevenson 1991). As mentioned
above, we have not explored the BD domain and we stopped the calculations
at 10 Gyrs for MS stars. Cooler models require non-grey atmospheres with grains.

\begin{figure}
%\picplace{2.5cm}
\epsfxsize=88mm
\epsfysize=80mm
\epsfbox{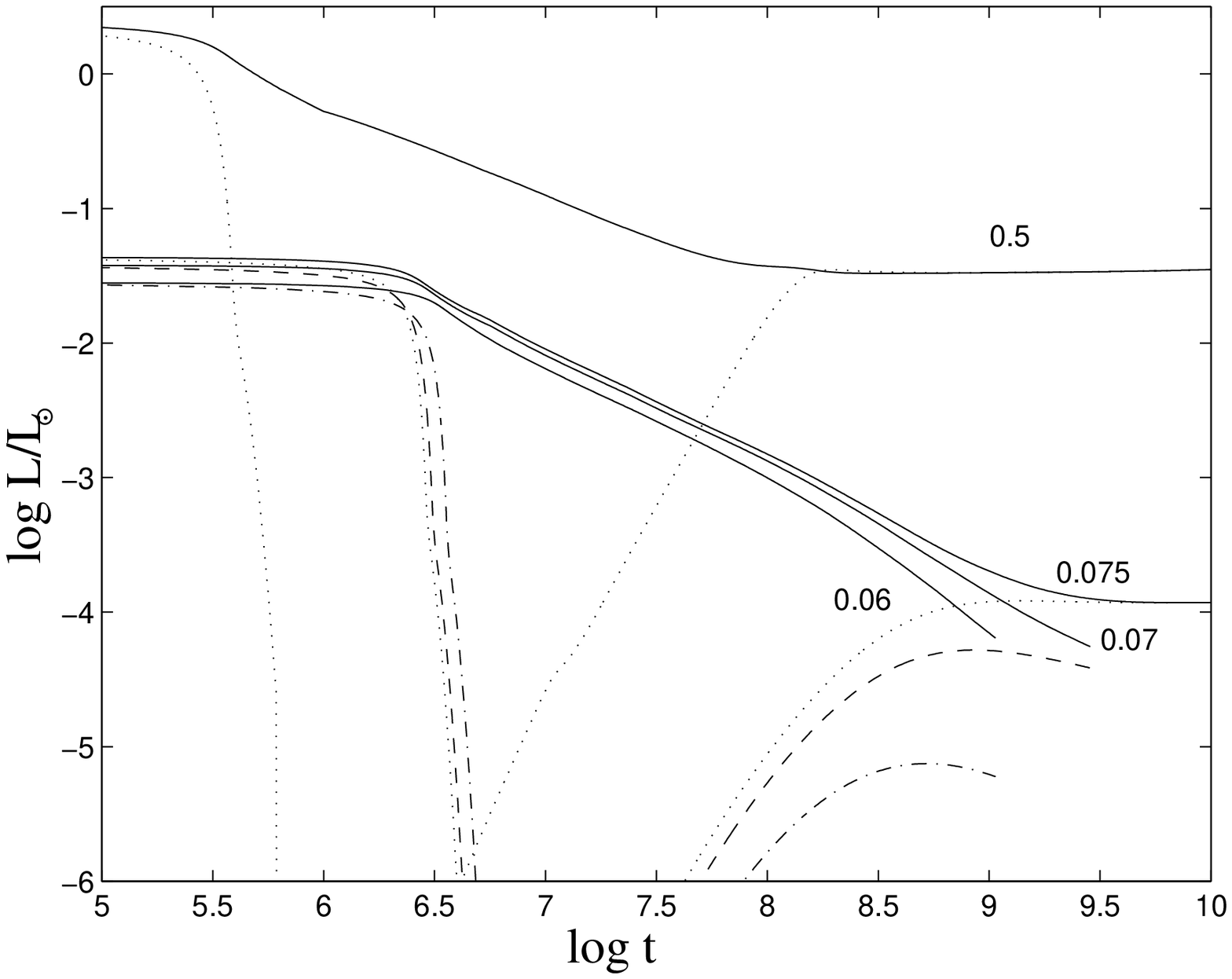} 
\epsfxsize=88mm
\epsfysize=80mm
\epsfbox{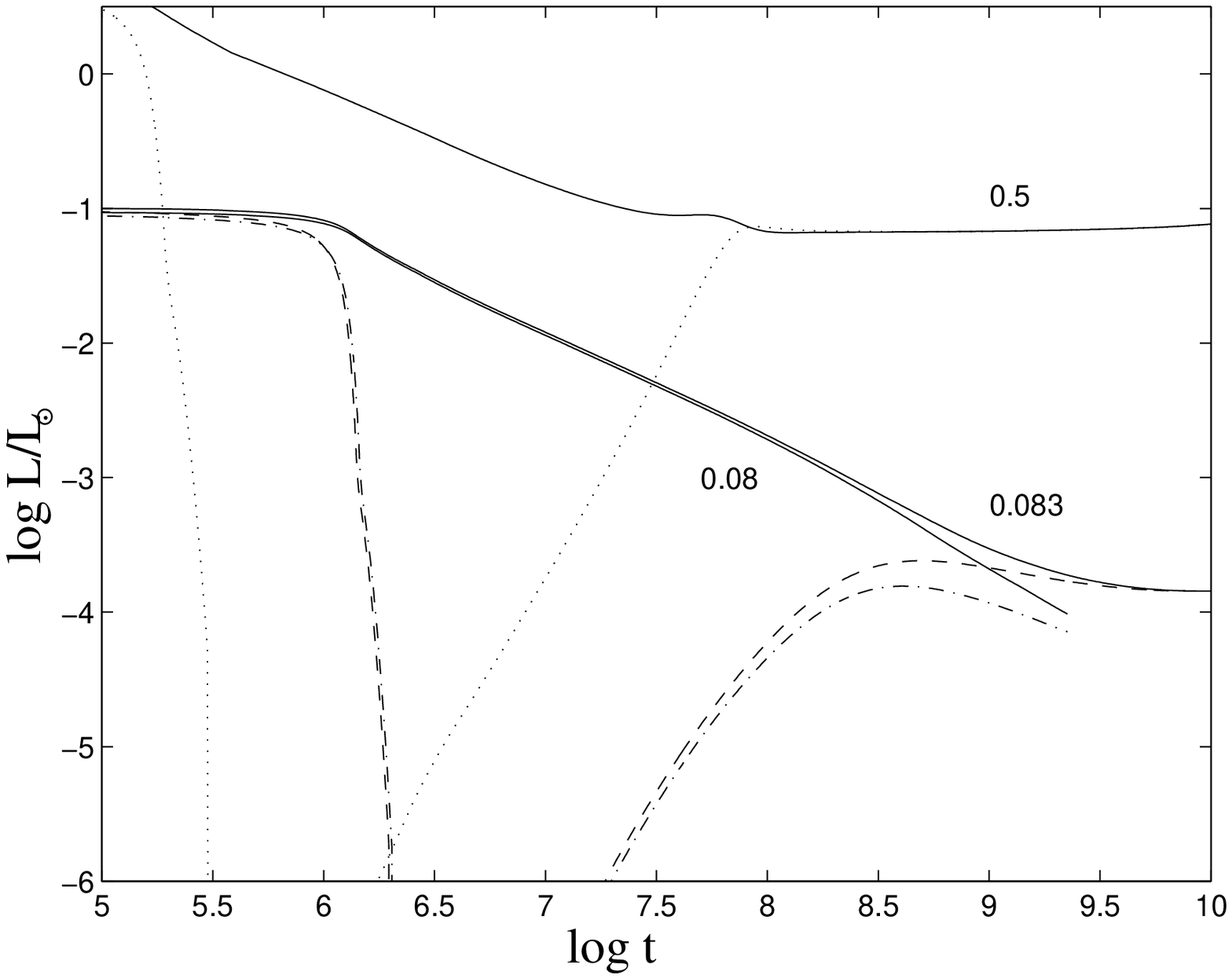}
\caption[ ]{{\bf (a)} Evolution of the luminosity for different
masses for $\mh=0$. Solid lines: total luminosity. Nuclear
luminosity $L_{nuc}$ : dotted line : 0.5 $\msol$ and 0.075 $\mso$;
dashed line :
$0.07 \msol$;
dash-dot line : $0.06 \msol$. {\bf (b)} Same as figure 8a for $\mh=-1.5.$
Nuclear luminosity $L_{nuc}$ : dotted line : 0.5 $\msol$;
dashed line :
$0.083 \msol$;
dash-dot line : $0.08 \msol$.}
\end{figure}
As already shown by Chabrier et al. (1996), stellar models based on
non-grey model atmospheres yield smaller HBMM than grey models, a direct
consequence of the lower effective temperature and luminosity, as discussed in the previous
section.
The larger the luminosity, for a given mass, the larger the 
required central temperature to reach thermal equilibrium, which in turns
implies a larger contraction (density) and degeneracy.
%Note that for solar metallicity, stars near the HBMM have
%$\te < 2200$ K. According to Tsuji et al. (1996), grain formation may
%alter the evolution at these temperatures. Work in this direction is under
%progress.

As illustrated on Fig. 8,
slightly below $\sim 0.072\,\msol$ (resp. 0.083 $\msol$)  for [M/H]=0 (resp. [M/H] $\le$ -1),
nuclear ignition still takes place in the central part of the star,
but cannot balance steadily the ongoing gravitational contraction. This defines the
{\it massive} brown dwarfs. The evolution equation thus reads
$L =\int \epsilon dm\,\,-\,\, \int T{dS\over dt} dm$, where the second term on the
right hand side of the equation stems from the contraction energy
plus the internal energy released along evolution. As shown in Fig. 7,
contraction is fairly small after $\sim 10^8$ yr, so that most of the
luminosity arises from the thermal content.
%The contributions from both sources (nuclear+contraction) are comparable
%(none of then is negligible)
%We elected to define this type of object as {\it transition object}.
%This is, however,
%a purely semantic point of view and some readers may prefer to call them brown dwarfs since these objects, though experiencing
Below about $0.07\,\msol$ (resp. $0.08\,\msol$) for [M/H]=0
(resp. [M/H] $\le$ -0.5), the energetic contribution
arising from hydrogen-burning, though
still present for the most massive objects, is {\it order of magnitudes} smaller than the internal energy, which provides essentially
{\it all} the energy of the star ($\epsilon << T|{dS\over dt}|$).
%We define these objects as genuine {\it brown dwarfs}. These three mass-ranges
%might be of some importance when discussing BD candidates.

As seen in the figures, objects with lower metallicity evolve at larger luminosities and effective temperatures, a well-known result.
The effects of metallicity on the atmosphere structure have been discussed extensively by Brett (1995) and AH95 but can be apprehended with intuitive
arguments. The lower the metallicity, the lower the opacity and the more transparent the atmosphere. The
same optical depth thus lies at deeper levels, i.e. at higher
pressure 
(${dP\over d\tau} = {g \over \kappa} $).
Therefore, for a given mass ($\log g$), the $(T,P)$ {\it interior} profile
matches, for a given optical depth $\tau$, an {\it atmosphere} profile with larger $\te$ (see Fig. 5).
This yields a larger luminosity, since the radius
barely depends on the atmosphere, as shown previously. 
%On the other hand,
%higher pressure and thus higher density favors convection ($F_{conv}\propto \rho$), which lowers the temperature gradient in the atmosphere. Therefore,
%at fixed $\te$ and $\log g$, a decreasing
%metal abundance yields a flatter temperature gradient, because of
%the more efficient convective transport at fixed optical depth (see e.g. AH95;
%Brett, 1995).
%This can be understood intuitively by considering the reverse case:
%if abundance of metals increases, the opacity per gram of matter %increases, therefore a decrease of the pressure and then of the
%density of matter, as well as an increase of the temperature gradient, %will help the photons to escape.
%Moreover, for a fixed abundance and $\log g$, the temperature gradient in %the atmosphere flattens as $\te$ decreases, due as above to an increase of %the convection efficiency (see AH 1995; Brett, 1995).
 %This will translate in different HBMM, for
%different metallicities, as discussed previously.

Objects above 0.15 $\msol$ reach the MS in
$t_{MS} \wig < 3 \times 10^8$ yrs, while it takes $3 \times 10^8 < t_{MS} < 2.5
\times 10^9$ yrs (resp. $t_{MS}\wig < 2\times 10^9$ yrs) for $\mh=0$ (resp.
$\mh \le -1.0$) for objects in the range $0.075 < M/\msol < 0.15$.
Objects above $M \sim 0.9 \mso$ (resp. $M \sim 0.8 \mso$) for $\mh=0$ 
(resp. $\mh \le -1$) start evolving off the MS at $t=10$ Gyr.
%(resp. $M=$) for $\mh=0$
%(resp. $\mh=-2.0$) start evolving off the MS after $t=10$ Gyr.

%Figure 9 shows the luminosity as a function of time for a 0.075 $\msol$ of %solar metallicity
%when assuming instantaneous mixing for the deuterium (see \S 2.3).
%As discussed in \S 2.3, this yields an overestimation of the d-abundance, %and then nuclear energy
%at a given central temperature. Thermal equilibrium can then be chieved %at a higher central
%temperature and lower density, reducing the level of degeneracy in the %star. As will be seen below, this strongly increases the luminosity  of %the star.

As shown on Figure 8, an object on the pre-MS contraction phase which will eventually become a H-burning
{\it star} ($M\wig > 0.075\,\msol$) can have the same luminosity and effective temperature as a bona-fide {\it brown dwarf}.

\subsection{Full convection limit}

Solar-like stars are essentially radiative, except for a small
convective region in the outermost part of the envelope, due to hydrogen partial ionisation,
and sometimes for a small convective core where nuclear burning takes place. As the mass decreases, the internal
temperature decreases ($T\propto M$), the inner
radiative region shrinks and
vanishes eventually for a certain mass
below which the star becomes entirely convective (see e.g. D'Antona
\& Mazzitelli 1985, Dorman et al. 1989).
This transition mass $M_{conv}$ 
has been determined with {\it grey} atmosphere models, which become invalid below $\sim 5000$ K, and thus must be
recalculated accurately.
%The transition from a fully convective structure to a structure with a %radiative core and a convective enveloppe
%depends on the interior temperatures, densities and opacities.
Figure 9 shows the interior structure of
stars with $M \ge 0.4 \msol$ as a function of time for [M/H]=0.
The pre-MS contraction phase
proceeds at constant $T/\rho^{1/3}$, i.e. constant $R_{OPAL}$ (
$R_{OPAL} = \log \rho / T_6^3$).
After $\sim 10^7$ years, a radiative core develops and grows. The physical
reason is
the decreasing
opacity after the last bump due to metal absorption (mainly $Fe$,), for
$T_6\wig >  2-3$ (this temperature decreases with
metallicity) 
(cf. Rogers and Iglesias 1992, Fig. 2). The radiative core thus appears
earlier for the more massive, hotter, stars.
The minimum mass
for the onset of radiation in the core is found to be
$M_{conv}= 0.35\, \msol$,  for all the studied metallicites
($-2\le \mh \le 0)$.
After $\sim 10^8$ years, the star reaches thermal equilibrium, nuclear fusion proceeds at the center,
and a small convective core develops for a certain time,
depending on the mass and the metallicity, bracketting the central
radiative region between two convective zones. 
We verified that the growth of the central convective core is governed by the
$^3He+^3He$ reaction. The nuclear energy released by the $p+p$ and
$p+d$ reactions is insufficient to generate convective instability.
As long as the reaction given by eqn.(2) dominates, the $^3He$ abundance,
and thus the convective core, increases. The situation reverses as soon as
the central temperature is high enough for $^3He$ to reach its
equilibrium abundance (eqns. (2)+(3)), wich decreases with increasing temperature (see e.g.
Clayton 1968,
Fig. 5.4). 
%%for the $^3He-^3He$ {\it destruction}
%reaction to become dominant and  for $^3He$ to reach its equilibrium value.
\begin{figure}
%\picplace{2.5cm}
\epsfxsize=88mm
\epsfysize=80mm
\epsfbox{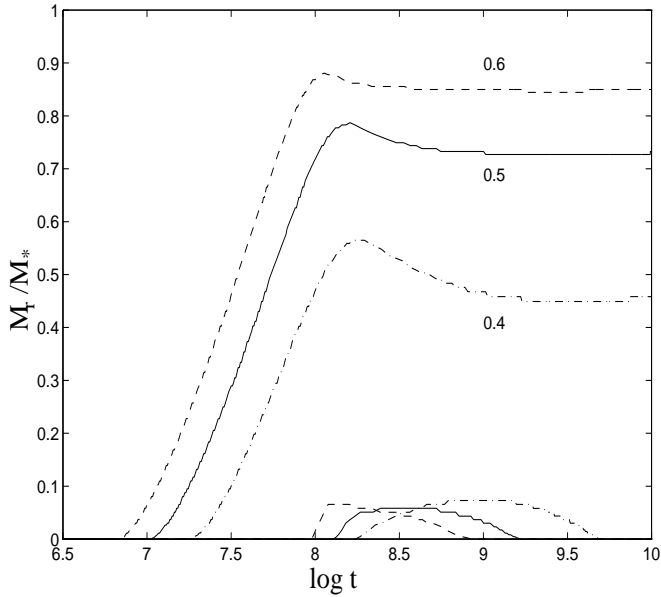} % where you want to insert a vbox for a figure
\caption[ ]{Evolution of the radiative zone
$M_{rad}/M_\star$
for [M/H]=0. The upper
curves determine the bottom of the convective enveloppe and
the lower curves the top of the convective core for 0.4 $\mso$ (dash-dot),
0.5 $\msol$ (solid line) and 0.6 $\mso$ (dash).}
\end{figure}

The extension of the afore-mentioned radiative
region decreases with temperature, and thus with mass, as shown on the figure.
For the afore-mentioned limit-mass $0.35\,\msol$, this inner radiative zone remains only for $\mh=-1.0$.
For all other (greater and smaller) metallicities, it vanishes as soon as the convective core appears. In this case
the 0.35 $\mso$ star will become fully
convective again after $\sim 3 \,10^9$ yrs.
This rather complicated dependence on metallicity
stems from
the subtle competiton between the {\it decreasing opacity},
which favors radiation, and the {\it increasing pressure and luminosity} 
which inhibate radiation and favor convection, with decreasing metallicity
($\nabla_{rad} \propto {L\kappa P \over T^4}$). This yields the minimum value for $M_{conv}$
for $\mh=-1$.

Table 1 gives the position of the bottom of the convective enveloppe
as a function of mass and metallicity. We also give
the results for the masses of the binary system YY-Gem.
Note that grey models, which have higher luminosity (see \S 4.1), thus have
a larger $\nabla_{rad}$, which favors convection and thus yields larger
convective envelopes and larger convective cores and
$M_{conv}$.
As will be shown below, the
onset of a radiative core, and the retraction of the bottom of the convective zone to outer, {\it cooler} regions bears
important consequences on the abundance of light elements in the envelope.
%For comparison, models based on grey-atmosphere models
%(using the Alexander Fergusson 1995 Rosseland means) yield M$_{conv}/\msol=$ ?, ? and ? respectively.

\subsection{Abundance of light elements}

 Observation of lithium in the atmosphere of VLMS is a powerful diagnostic for
the identification of
genuine brown dwarfs, as proposed initially by Rebolo and collaborators (Rebolo, Martin \& Magazz\`u 1992). The first theoretical analysis of light-element burning in VLMS, and the expected abundances along
evolution, were carried  by Pozio (1991) and Nelson, Rappaport and Chiang (1993). As for the convective limit, this analysis must be
redone with updated EOS, screening factors and atmosphere models.
The initial abundances were taken to be $[^7Li]_0=10^{-9}$, $[^9Be]_0=10^{-10}$
and $[^{11}B]_0=3\times 10^{-11}$, as in Nelson et al. (1993). The modification of the abundances along
evolution, i.e. the depletion factor, is given by $X_i/X_{i_0}$, where $X_i$ is the abundance of element $i$ at a given time and $X_{i_0}$ is the
afore-mentioned initial abundance. The burning temperatures for these elements (in the vacuum)
are $T_{Li}\sim 2 \times 10 ^6$ K,
$T_{Be}\sim  3 \times 10^6$ K and $T_{B}\sim 4 \times 10^6$ K, respectively.

As for hydrogen, burning ignition temperatures
translate into minimum burning masses.
Using the ion+electron screening factors mentioned in \S2.2, we get the following values, for solar metallicity :
$M_{Li}/\msol = \, 0.055$, $M_{Be}/\msol = \, 0.065$
and $M_{B}/\msol = 0.08$. For comparison, Nelson et al. (1993) find that $\sim \, 50\%$
of $^7$Li is burned in a 0.059 $\mso$, whereas this value is already reached in our 0.055 $\msol$ model. D'Antona \& Mazzitelli (1994)
obtain $M_{Li}/\msol = \, 0.065$. 
%***what about central temp. ?
This less efficient nuclear burning
in Nelson et al. and DM94
stems on one hand from the grey approximation, which yields larger $L$ and $\te$
and thus central densities, which favors the onset of degeneracy (cf. \S 3.1.2
and Fig. 10 below), and also
from the smaller
Graboske et al. (1973) screening factors (see Figure 3).

%our larger screening factors (see Figure 3). 
%For Be, 50\% is consumed in a 0.075 $\msol$ and we find $\sim 0.07 \msol$. %For $^{11}$B, they
%obtain 50\% of depletion in a 0.092 $\msol$ star in a hubble time and
% we obtain already in a 0.09 $\mso$ a depletion factor of 6.
These effects, and the metallicity dependence, are illustrated in Figures 10a-b
which display the evolution of central temperature and lithium-abundance, respectively,
for different metallicities.  Comparison is made with a solar metallicity
model of DM94 (cf. their Table 7).
The {\it denser} metal poor stars (and grey models) reach the limit of degeneracy ealier ($\psi \propto \rho^{-2/3}$), yielding a lower
maximum central temperature for metal poor objects.
The direct consequence is an increasing minimum burning mass with
decreasing metallicity,
$M_{Li}/\msol = \, 0.06$ for [M/H] $\le -1$ instead of
$0.055$ for [M/H] $\sim 0$.
%The same arguments apply to the $\NextGen$ models with
%solar metallicty which predicts a higher luminosity than the $Base$models.
%However, the $\NeG$ and grey models based on the Krisna-Swamy
%$T - \tau$ relationship give a quite similar evolution in the brown-dwarf
%regime. Whereas the structure of the solar models with different %atmosphere models is quite similar, the situation is different
\begin{figure}
%\picplace{2.5cm}
\epsfxsize=88mm
\epsfysize=80mm
\epsfbox{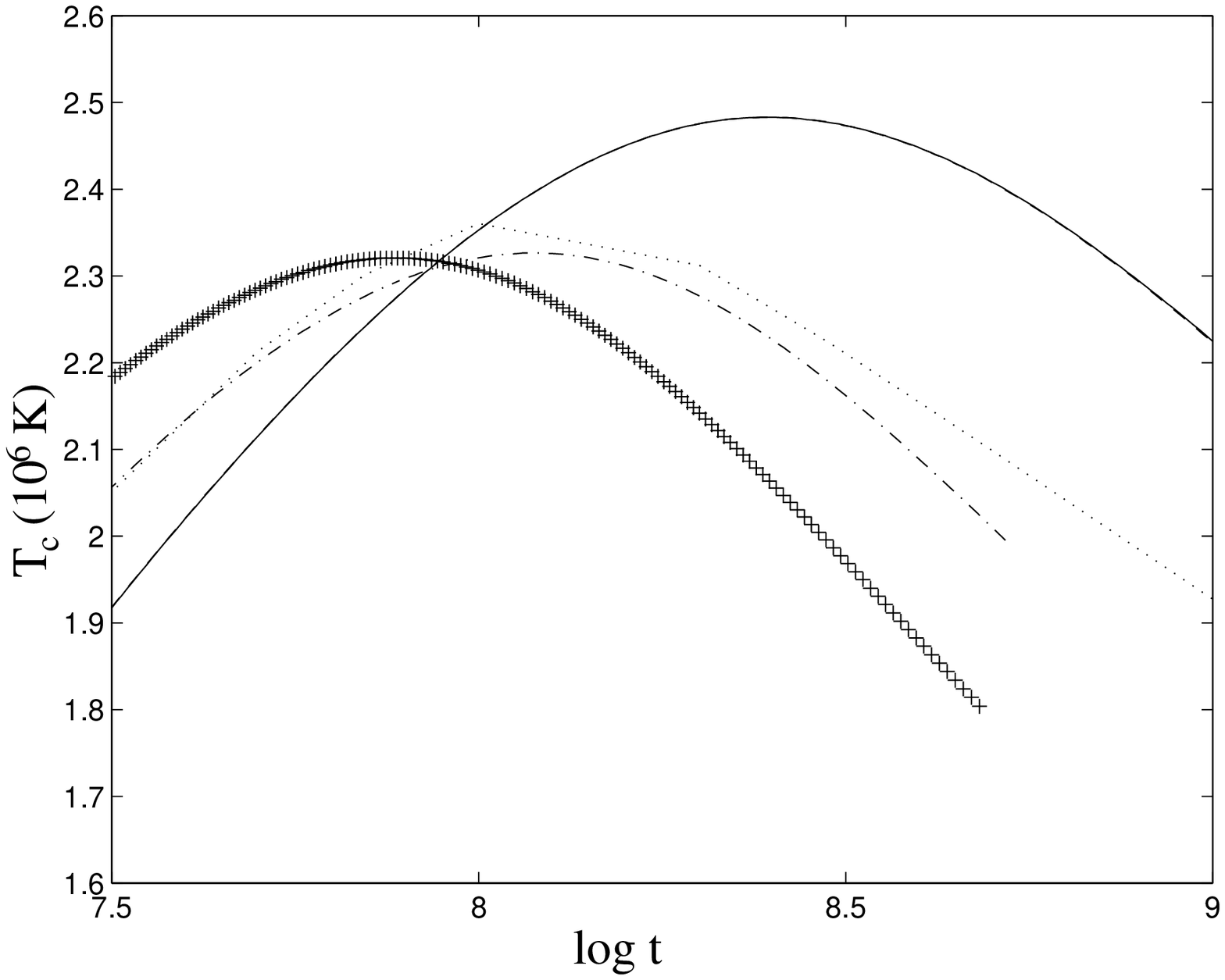} % where you want to insert a vbox for a figure
\epsfxsize=88mm
\epsfysize=80mm
\epsfbox{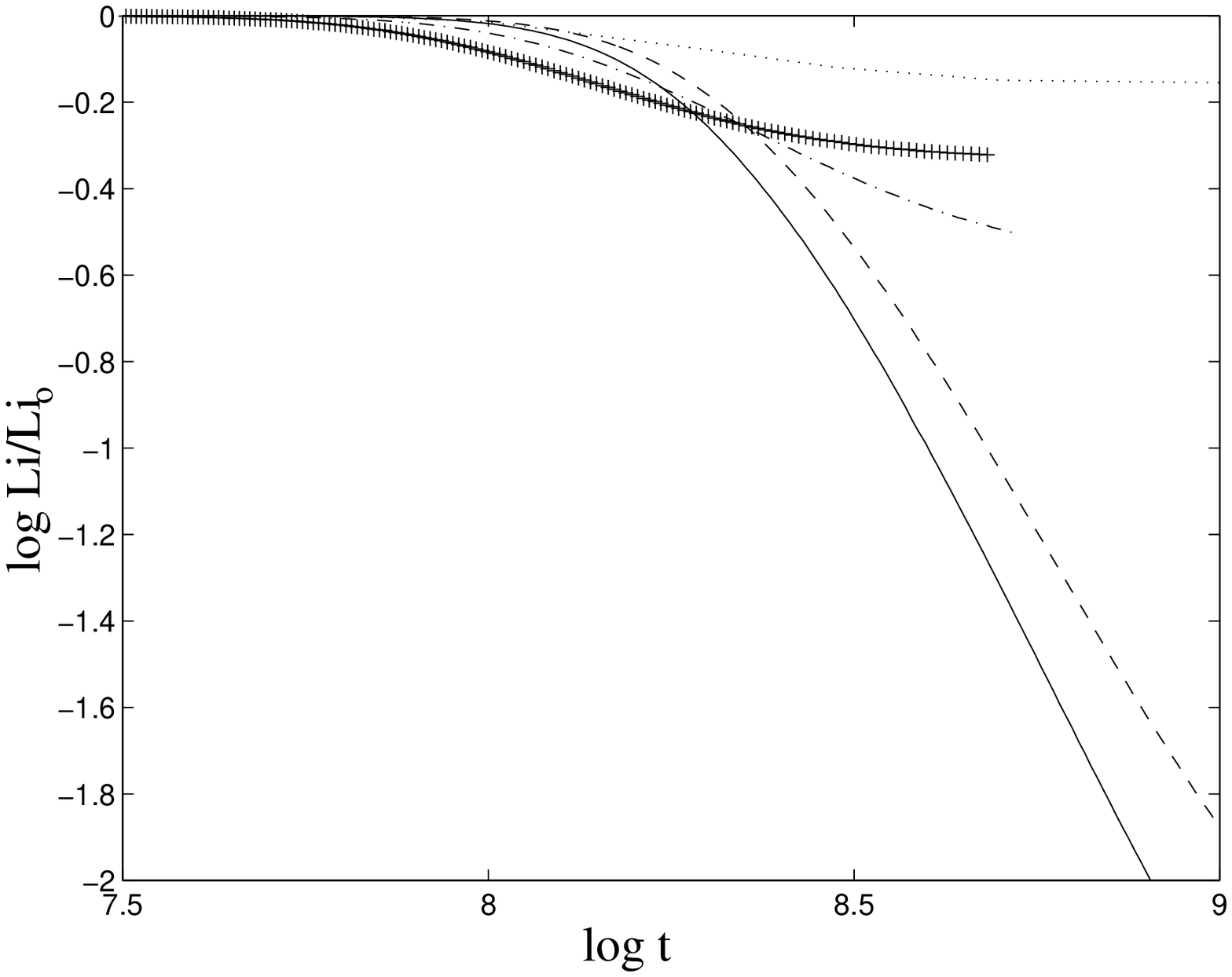}
\caption[ ]{{\bf (a) } Evolution of the central
temperature in a 0.06 $\msol$ brown
dwarf. Solid line : $\mh=0$; dot-dash line : $\mh=-1.0$; ($+$) : $\mh=-2.0$.
Dotted line: 0.06 $\msol$ of DM94 for [M/H]=0.
Dashed line : calculations with the Graboske et al. (1973) screening factors,
for $\mh=0$.
Note that the
solid and dashed curves are undistinguishable, and illustrate the negligible
effect of the electronic screening factor on the {\it evolution} of LMS.
{\bf (b)} Same as figure 10a for the abundance of Lithium
w.r.t. its initial abundance. Same legends as in figure 10a.}
\end{figure}

%The same arguments apply to the $M_{BeBMM}$, as it concerns as well the BD %regime. Interestingly enough, we find the same dependance on the
%metallicity for the B burning efficiency. Indeed, in a 0.09 $\mso$, we
%find that the depletion factor for [M/H]=0 is 6, as already mentioned, %compared
%to 3.2, 2.8 and 1.5 for respectively [M/H]=-1, -1.5 and -2. This stems %also
%from the increasing importance of degenaracy as [M/H] decreases and below
%0.1 $\mso$. As will be illustrated in the next section (\S 4.1), stars
%below this mass limit burn their hydrogen on the MS at {\it lower} central %temperature as the metallicity decreases because of higher density and
%level of degeneracy.

Figure 11 shows the abundances of $Li$, $Be$ and $B$ as a function of {\it mass}
and metallicity.
The gaps correspond to fully convective interiors,
as described in \S3.2.
In that case, convective mixing brings the elements
present in the envelope down to the central burning region where they are
destroyed.
Above 0.4 $\msol$, the central radiative core appears and the bottom
of the convection zone retracts to cooler regions, as described
previously. As mentioned above, this depends mainly on the
central opacity : the larger the opacity, and thus the metal-abundance,
the larger the central temperature required to allow radiative transport ($\nabla_{rad}
\propto \kappa/T^4$).
This yields more efficient depletion with increasing metallicity,
as illustrated on the figure.
%As shown on this diagram, the observation of Lithium absorption features in an %object
%{\it older than $t\approx ?$ years} will unambiguously identify it as a %bona-fide brwon dwarf.
%For  stars which develop a radiative core, the depletion of light elements
%highly depends on the central temperature at which it forms. As already %mentionned in the previous section, the appearance of radiation in the %core depends mostly on the central opacity. Consequently, . This effect is %illustrated in Fig. 15 where the evolution of the bottom of the convective %envelope in stars of different metallicity is displayed along the central %temperature. For any masses M $\ge 0.4 \mso$, the central temperature at %which convection regresses is higher in the solar case.  The direct %consequence is a more efficient depletion. Thus,
Lithium is totally destroyed in the mass range 0.075-0.6 $\mso$ for [M/H]=0, whereas it reappears for M $\wig > 0.5 \mso$ for
[M/H] $\le$ -1. However, below [M/H]=-1.5, the situation seems to reverse, as
the [M/H]=-2 case shows a slightly higher level of depletion.
At such low metallicities, the dependence of the opacity on the metallicity for T$_c \ge 3\times 10^6 K$ decreases rapidly and the dominant effect is now the higher luminosity at [M/H]=-2, which implies a larger
central temperature to favor radiation ($\nabla_{rad}\propto L/T^4$).
%***
The depletion factors in the stellar interiors are given in Tables 2-7. The lithium depletion factors in the brown dwarf regime are given
in Chabrier et al. (1996).
A complete description of light elements depletion
in this regime, which implies evolutionary models based on dusty atmosphere models, is under progress.

\begin{figure}
%\picplace{2.5cm}
\epsfxsize=88mm
\epsfysize=90mm
\epsfbox{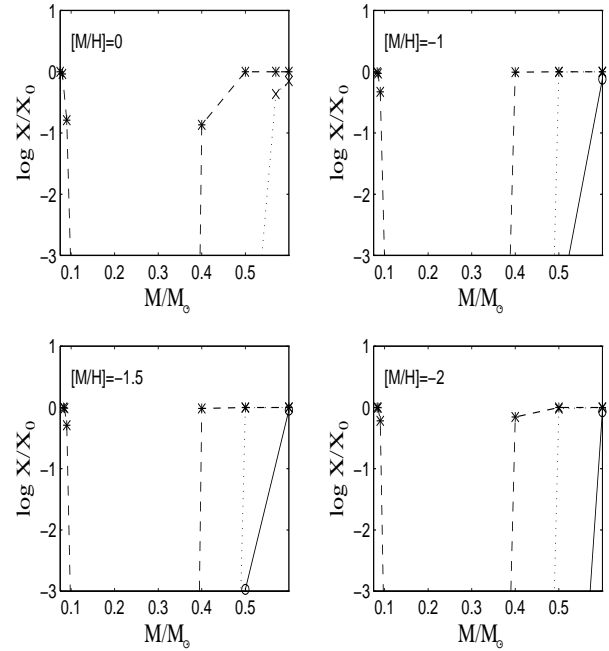} % where you want to insert a vbox for a figure
\caption[ ]{Abundances of light elements as a function of mass and
metallicity
at $t=$10 Gyrs. Only results in {\it stars}
are shown ( $M \ge 0.075 \mso$ for [M/H]=0 and $M \ge 0.083 \mso$ for [M/H] $\le $ -1).
Abundances normalized to their initial value are shown for Li (solid,$O$), Be (dot,$X$) and B(dash, $\star$).}
\end{figure}

\section{Mass-dependence of the photospheric quantities}

\subsection{Mass-luminosity relationship}

Figure 12 shows the mass-luminosity (ML) relationship for VLMS for three metallicities, $Z=Z_\odot$, $Z=10^{-1.5}\times Z_\odot$ and $Z=0$, for
$t=$10 Gyrs.
%For [M/H]=0 and -1.5, the $\NeG$ and $base$ models are shown, the lower %luminosity
%obtained with the $base$ atmosphere models for the solar case resulting mainly %from the overestimation
%of the opacity due to the Straight Mean method (cf. Chabrier et al. 1996).
%Differences between  $\NeG$ and $base$ models decreases with the
%metallicity,
%and both sets of models give quite similar results for [M/H]=-1.5,
%the discrepancy observed however below 0.1 $\mso$ is due to a different helium %abundance (Y=0.275 in Baraffe et al. 1995 and Y=0.25 in the present work).
%Near the HBBM, the structure of stars is indeed very sensitive and a slight %increase of Y leads to a subsequent increase of L. Fig. 16
%shows the resultas obtained for Z=0 with .
The zero-metal \footnote{
The $Z=0$ non-grey atmosphere models were kindly provided
by D. Saumon.} case sets the upper limit for the luminosity
for a given mass.

We first note the well-known wavy behaviour of the ML
relation (see e.g. DM94).
The change of slope below $M \sim 0.4 - 0.5\, \msol$ is due to the formation of H$_2$
molecules in the atmosphere (Auman 1969; Kroupa, Tout \& Gilmore, 1990), which occurs at higher $\te$ for decreasing metallicity, because of
the denser
atmosphere (see SCVH).
%For example, a $H_2$ number fraction $x_{H_2}\sim$ 25\% is obtained for
%$\te \sim  3600$ K, $\rho\sim 10^{-6}$ g cm$^{-3}$ in a 0.5 $\mso$ with [M/H]=0,
%whereas the same fraction is reached for $\te \sim  ??$ K, $\rho\sim 10^{-??}$ g cm$^{-3}$ for the same mass with [M/H]=-1.5.
 The steepening of the ML relation near the lower-mass end reflects the
onset of ongoing degeneracy in the stellar interior, as demonstrated in \S 3.1.1. 
The previous stellar models of BCAH95, based on the $Base$ AH95 atmosphere models,
are shown for comparison. We note that for solar-metallicity, the substantial
improvement in the most recent atmosphere models translate into significant
differences ($\sim 15\% - 25\%$ in $L$ and $\sim 200$ K in $\te$), the
Straight-Mean approximation used in the
$Base$ models leading to an overall
overestimated opacity. This
difference between the models vanishes for lower metallicities
(see Allard et al., 1997b). 
Note that the present models have been shown to reproduce accurately the mass-$M_V$ and mass-$M_K$
relationship determined observationaly by Henry \& Mc Carthy (1993)
down to the bottom of the MS (Chabrier et al., 1996; Allard et al., 1997a).
The $ML$ relations for different metallicities are given in Tables 2-7.
The
comparison with other models/approximations is devoted to the next subsection.
\begin{figure}
%\picplace{2.5cm}
\epsfxsize=88mm
\epsfysize=100mm
\epsfbox{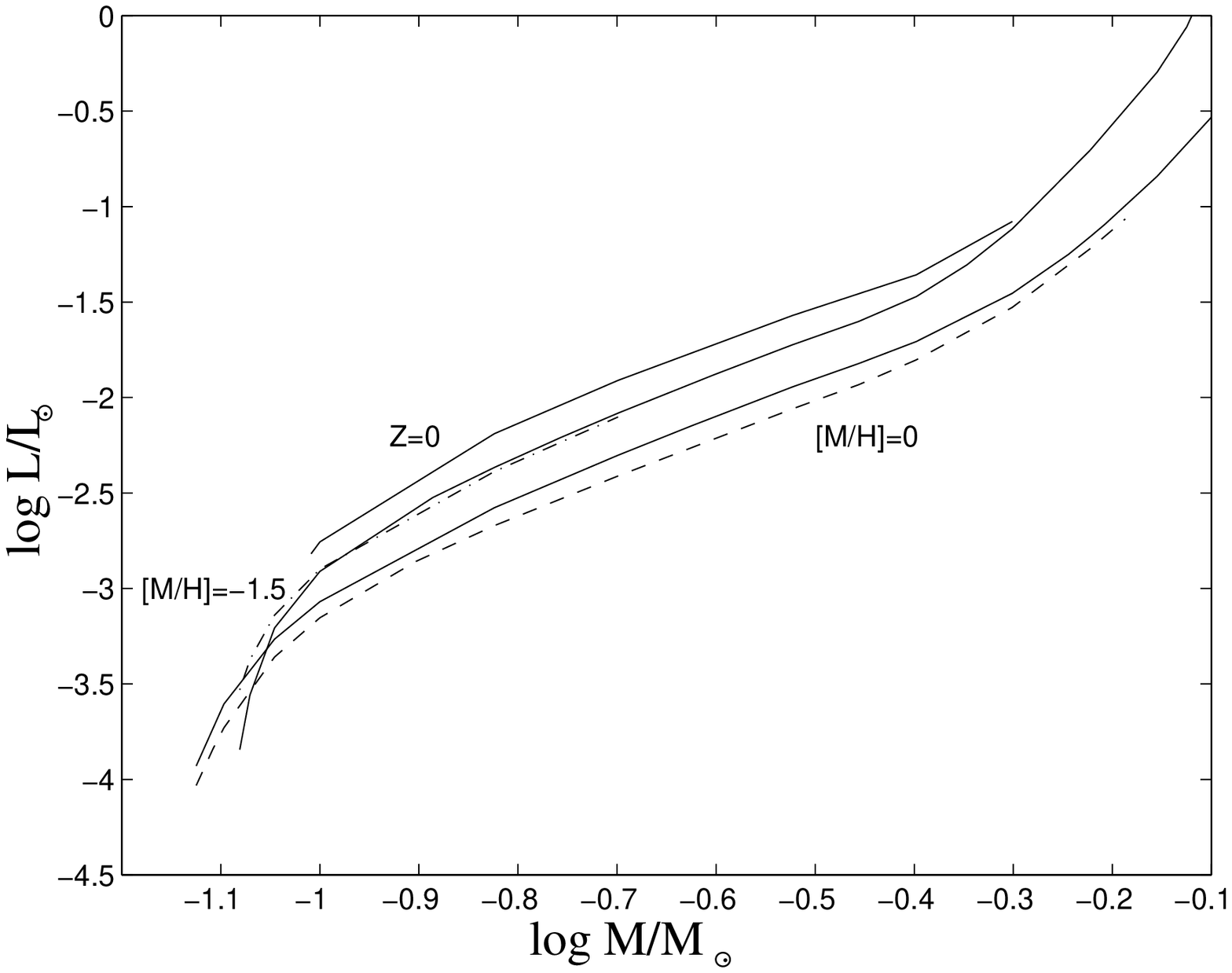} % where you want to insert a vbox for a figure
\caption[ ]{Mass-luminosity relationship for
[M/H]=0, $\mh=$-1.5 and the zero-metal
case Z=0. The previous results of Baraffe et al. (1995) based on the $Base$ atmosphere models are also shown ( dashed curve for [M/H]=0 and dash-dotted curve for [M/H]=-1.5).}
\end{figure}

\subsection{Mass-effective temperature relationship}

Figures 13 display the mass-effective temperature relationships for sub-solar
(Fig. 13a) and solar (Fig. 13b) metallicities.
A shown in Fig. 13a, our zero-metal models reproduce correctly the models of Saumon et al. (1994)(filled circles).
Figures 13 also display the $m-\te$ relation
obtained  
with the Krishna-Swamy $T(\tau)$ relationship,
with (dotted line) and without (dot-dashed line) 
convection in the optically thin region (cf. \S 2.5).
The figures clearly show the overestimated effective temperature obtained
by grey models using a diffusion approximation (Burrows et al. 1989, ($+$); DM94 (squares); DM96 (triangles)),
as demonstrated in \S 2.5.

As already mentioned, the Krishna-Swamy relationship with convection {\it arbitrarily suppressed} at $\tau <1$ (Alexander et al. 1996, ($X$)) leads to
less severe discrepancy in the region where convection {\it does} penetrate into the optically thin layers (2500 K $< \te  <$ 5000 K). This paradoxal and inconsistent situation clearly illustrates the dubious reliability of such a treatment,
and reflects the unreliable representation of the effects of atmospheric convection within a grey-approximation. This is clearly illustrated by the rather unphysical atmospheric profile obtained within this approximation, as
shown in Figure 5b.
%For solar metallicity, however, convection in optically thin layers is less efficient ($F_{conv} \propto \rho$),
%both KS grey approximations yield the same atmosphere profiles (see Fig. 5a).
%is not determinant for the atmosphere structure (Fig. 5a; \S 2.5).
For solar metallicity, however, both KS grey approximations yield a similarly good match to the innermost
part of the atmosphere profile (see \S 2.5).
It is the reason why the
KS treatment {\it with} convection in the optically thin region, as used in
Dorman et al. (1989) (filled circles), yields a reasonable agreement at
solar metallicity,
whereas it yields
severe discrepancies for lower metallicities.
This reflects the significant overestimation of the convective flux as
density and pressure increase with decreasing metallicity.
%This reflects the important dependence
%of the convective flux on density (and thus on metallicity).
%This disagreement  at lower metallicity
%stems from an overestimation of the convective flux, due to higher densities,
%when using a $T(\tau$) relationship.
Models based on the Eddington approximation predict even higher
$\te$ at a given mass (see Fig. 5).

% Such surprising result may stem  from the following arguments:
% when convection is taken into account in the optically thin region, the
% calculation of the radiative temperature gradient is required in order
% to estimate the superadiabaticity. However, when using a T - $\tau$
% relationship, the real radiative temperature gradient is not accessible,
% as it needs to solve the transfer equations,
% and thus requires an approximation as  correction
%  prescribed by. Such approximation, though improving the situation
% compared to results obtained with the Eddington approximation (i.e. a
% diffusion approximation) still overestimates the total flux for a given
% T-P atmosphere profile and leads to higher $\te$ for a given mass.
The unreliability of any $T(\tau)$ relationship for VLMS becomes even
more severe
near the bottom of the MS ( $m< 0.1 \msol$), as shown on the
figures. They yield too steep $m-\te$ relationships in the stellar-to-substellar
transition region and thus too large HBMMs, by $\sim 10\%$,
as discussed in \S 3.1.2.
The difference between grey and non-grey calculations vanishes for $\te \wig > 6000$ K,
i.e. $\sim 0.8\,\msol$ for metal-depleted abundances.
For a
0.8 $\msol$ star with $\mh <0$,
the difference between models
based on non-grey AH97   atmospheres and on grey models calculated with Alexander and Fergusson (1994)
Rosseland opacities amounts to $\sim$ 1-2\% in $\te$ and less than 1\% in L.

%The present non-grey models, the HBBM is $\sim 0.07 \msol$
%for [M/H]=0,
%whereas Dorman et al. (1989) finds with the FGC EOS 0.08 $\msol$, Burrows et %al. (1993) find
%0.076 $\msol$. We also find that our grey models
%(Eddington or Krisna-Swamy) predict a HBBM $> 0.072 \msol$.
%We have also checked that the grain formation does not alter this
%picture by evolving stars around the HBBM with grey models based on dust-free
%Rossland opacities, kindly provided by D. Alexander (1995, priv. com) (cf. %Chabrier et al. 1996 and \S 2.5). We therefore stress out that the use of a
%T - $\tau$ relationship, even corrected to the diffusion approximation and
%grey effects as the Krishna-Swamy relationship, can indeed give results
%which departs slightly from models based on non-grey models {\it if}
%atmospheric convection is inefficient.
\begin{figure*}
%\picplace{2.5cm}
\epsfxsize=120mm
\epsfysize=100mm
\epsfbox{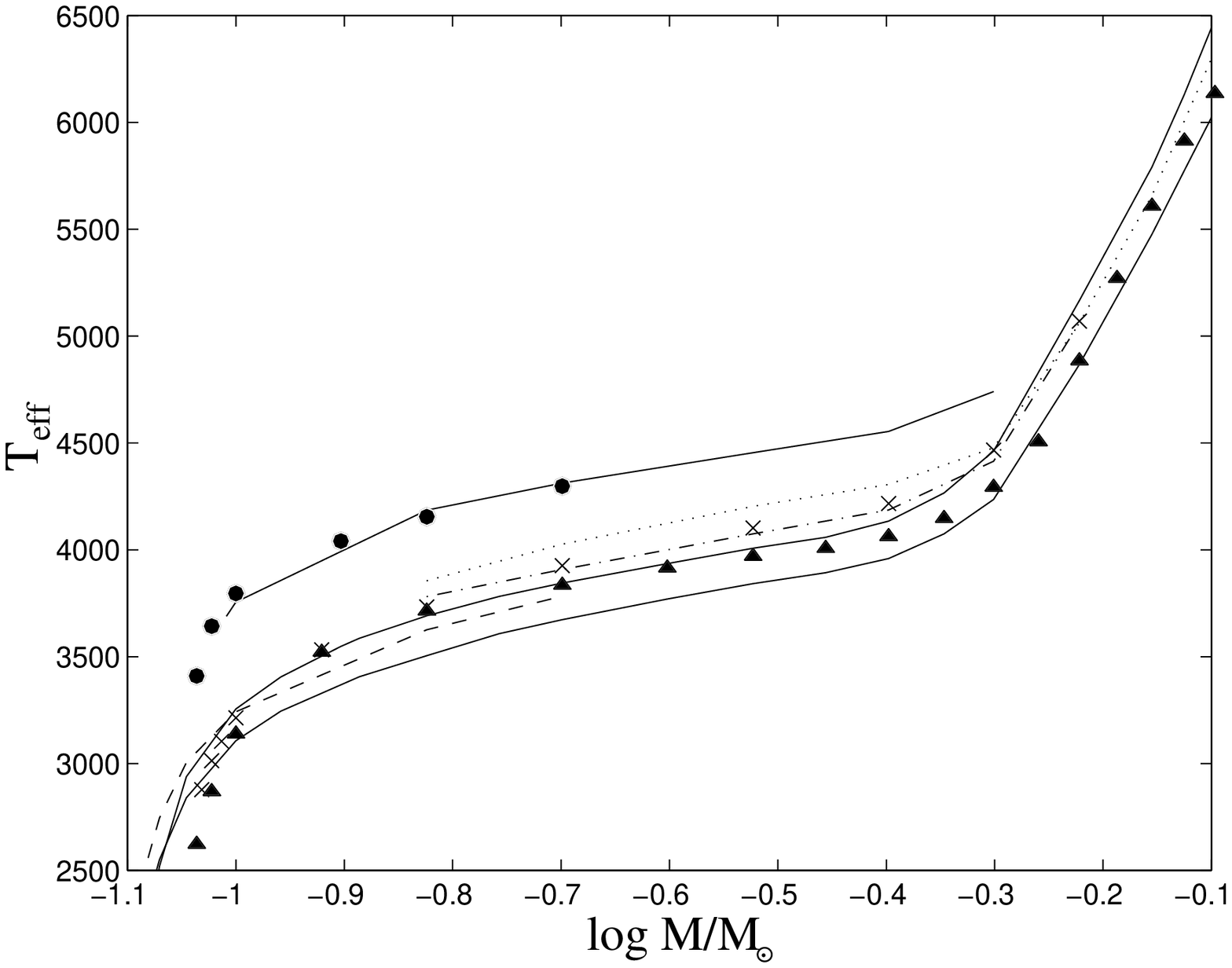} % where you want to insert a vbox for a figure
\epsfxsize=120mm
\epsfysize=110mm
\epsfbox{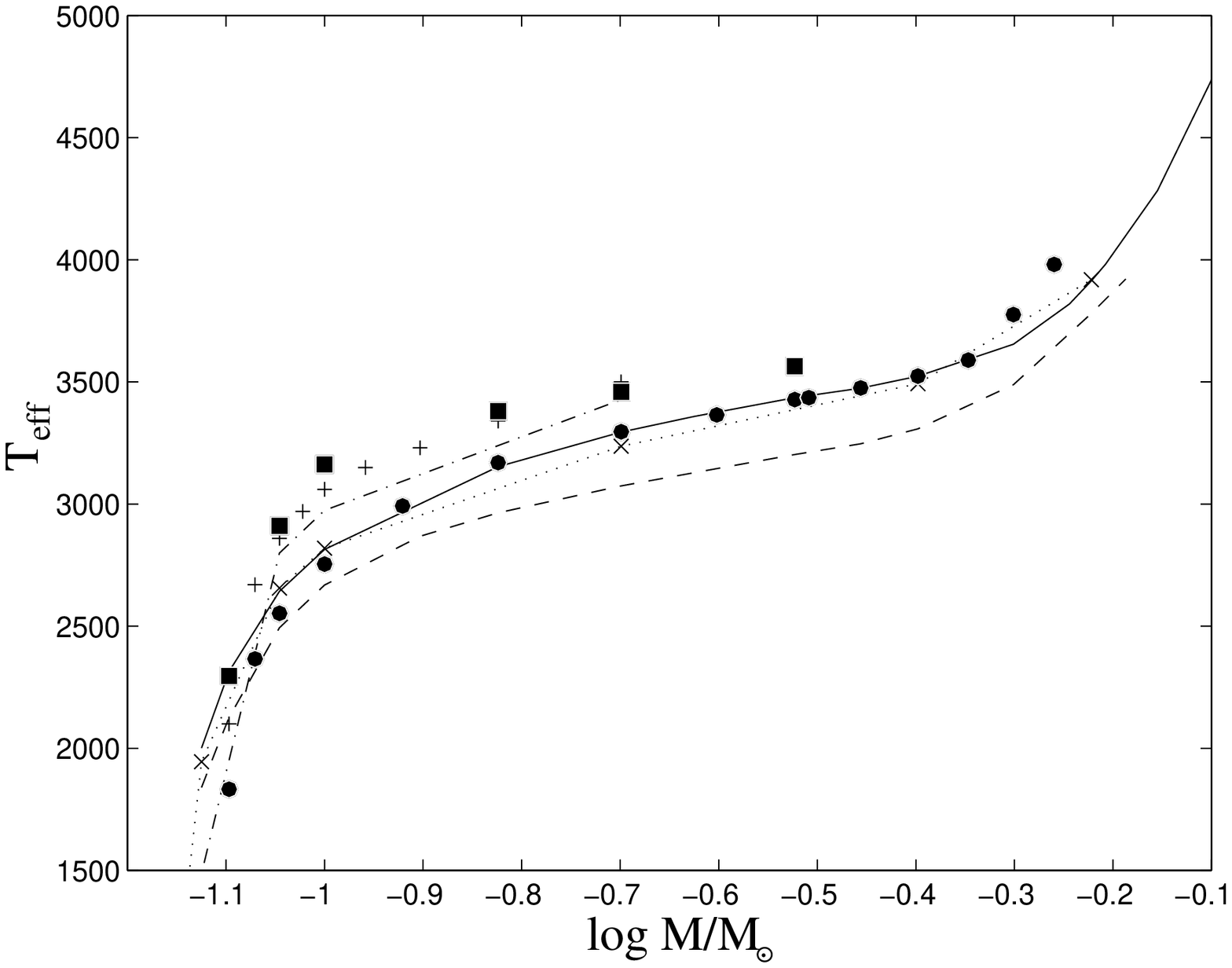}
\caption[ ]{{\bf (a) } Mass-effective temperature relationship for
low-metallicity.
Solid lines : present models for $Z=0$ and $\mh= -1.5,\, -1.0$, from top to bottom.
Dashed line : non-grey models based on the $Base$ atmosphere models for $\mh=-1.5$ (BCAH95).
Dotted line and dot-dashed line : models for [M/H]=-1.5 based on the
Krishna-Swamy $T(\tau)$
relationship {\it with} (dot) and {\it without} (dash-dot)
 convection in the optically thin region. Crosses : the [M/H]=-1.5 models of Alexander et al. (1996). Triangles : the [M/H]=-1.0 models of D'Antona and Mazzitelli
(1996). Full circles : the Z=0 models of Saumon et al. (1994). {\bf (b)} Same as figure 13a for
solar metallicity. Solid and dashed lines as in Fig. 13a.
Dash-dotted line : Eddington
approximation. Dotted line (and crosses) : Krishna-Swamy relationship.
Comparison
with previous works: Burrows et al. (1989), models G ($+$); D'Antona \&
Mazzitelli (1994), models with the Alexander opacities and the MLT
(full squares);
Dorman et al. (1989), models with the FGV EOS (full circles).}
\end{figure*}

%The increasing density, and gravity (see e.g. Fig 6)
%with decreasing metallicity
%favors convection in the atmosphere (Brett, 1995), and thus 
%flattens the temperature gradient and leads to higher $\te$ for a given mass %(cf. \S3.1.2).
The different mass-$\te$ relations are given in Tables 2-7.
Differences between effective
temperatures as a function of the metallicity, for a given mass,
$(\Delta \te/\Delta Z)_M$, decrease with mass. The 0.5 $\msol$ star with [M/H]=-1.5 is $\sim$ 800 K hotter than
its solar metallicity counterpart, whereas the difference reduces to $\sim 300 K$ for the
0.09 $\msol$. This stems from the decreasing sensitivity of the
atmosphere structure to metal abundance with decreasing $\te$ 
%interpreted as
%a saturation of the convection efficiency 
%with decreasing luminosity 
%and a decline of metal-line opacities
%with decreasing effective temperature 
(see e.g. Allard, 1990; Brett, 1995; AH97).

%*********************************************************
%
%**** Je n'y est pas touche. Que fait-on ? ***
%\subsection{CM Draconis and YY Geminorum}
%
%The eclipsing binary systems
%CM Daconis (M $\sim 0.2 \mso$) and YY Geminorum (M $\sim 0.6 \mso$) provide a
%stringent test for the structure and evolutionary models of LMS. The
%$NextGen$ atmosphere models have been shown to reproduce accurately the
%$\te$-color relation, in ?? (Allard et al. 1997), whereas
% the first generation of the present models
%(BCAH95) was shown to reproduce for the first time all the observed quantities %of these objects for the correct masses (Chabrier \& Baraffe, 1995).
%Figures 15 display the same analysis, i.e. the $m-R$ and $m-\te$
%relationships for the present models, for solar metallicity.
%As already mentioned, different atmosphere models
%hardly modify the $m-R$ relationship, which reflects essentially the accuracy
%of the EOS.
%As shown on figure 19b, the present models slightly overestimate the effective %temperature of $CM-Dra$ ($\te = 3150 \pm 100 K$).
%As mentioned in Chabrier and Baraffe (1995), the observed
%colors of CM Dra were best fitted by the $base$ models of metallicity
%[M/H] = -0.5 and $\te$ = 3300 K. This sub-solar metallicity, characteristic of
%the old-disk population, is consistent with the rather large space velocity of
%the system (see Lacy 1977) and the difference in the effective temperature is
%consistent with a $\sim 10\%$ error on the parallax.
%%
%*** bla, bla, bla, ca marche plus !!!! ***
%************************************************************

\section{Conclusion}

In this paper, we have presented new calculations aimed at describing the
structure and the evolution of low mass stars, from solar masses down to the
hydrogen burning limit, for a wide range of metallicities. These calculations include the most recent physics
aimed at describing the mechanical and thermal properties of these
objects - equation of state, screening factors for the nuclear reaction rates,
non-grey atmosphere models
-.
The Saumon-Chabrier hydrogen-helium EOS gives the
most accurate description of the internal properties of these objects
 over the entire afore-mentioned mass range. Given their negligible number abundance, metals play essentially
no role on the EOS itself and their presence is mimicked adequately by an
effective helium fraction. Of course they do play a role as electron donors
for the opacities and must be included in the appropriate ionization equilibrium
equations when {\it opacities} are concerned.
Note that, for the {\it densest} stars ($0.1\,\msol$, $\mh=-2.0$, $\log g=5.5$),
departure from ideality at $\tau =100$ (resp. $\tau=1$) in the atmosphere
is found to represent $\sim 3\%$ (resp. $\sim 1\%$) on the adiabatic gradient
(as obtained from a comparison between the complete SC EOS vs an ideal gas EOS). Thus, under LMS conditions,
an ideal (Saha) EOS can be safely used over the entire atmosphere, for metallicities
$\mh \ge -2$. Note that an (incorrect) grey approximation yield denser atmospheric profiles
(cf \S 2.4) and thus will overestimate non-ideal effects.

% In the atmosphere of LMS, this concerns
%the perfect gas domain and is easily solved by the Saha equations. 
We show
that under LMS conditions, the {\it responsive} electron background participates
to the screening of the nuclear reactions, and must be included for an accurate
determination of the various minimum burning masses and of the abundances of light
elements along evolution. We show that, near the bottom of the main sequence,
the deuterium lifetime against proton capture in the PPI chain is order of magnitudes smaller
than the convective mixing time. Instantaneous mixing is thus no longer satisfied.
This yields the presence of a deuterium
gradient in the burning core, which bears substantial consequences on the
determination of the luminosity near the brown dwarf regime.

We have examined carefully the effect of various grey-like approximations on the evolution
and the mass-calibration of LMS.
Under LMS conditions, these prescriptions are incorrect,
or at best unreliable, and yield inaccurate mass-luminosity and
mass-effective temperature relationships, which in turn yield inaccurate {\it mass functions} from
observed luminosity functions. We examine the behaviour of the
different stellar macroscopic quantities, radius, temperature, luminosity,
as functions of mass, time and metallicity. We link the general behaviour of
these quantities to intrinsic physical properties of stellar matter, in
particular the transition from classical to quantum objects and the onset of
convection or radiation in the stellar interior and atmosphere.
We derive new limits for the hydrogen-burning minimum mass, effective temperature
and luminosity, for each metallicity. These limits are smaller than the values determined previously,
a direct consequence of non-grey effects in the atmosphere.

We believe the present calculations to represent a significant improvement
in low-mass star theory and in our understanding of the properties of cool
and dense objects. This provides solid grounds to examine the structure and
the evolution of
substellar objects, brown dwarfs and exoplanets. 

At last the present models provide reliable mass-luminosity relationships,
a cornerstone for the derivation of accurate mass functions (M\'era, Chabrier \& Baraffe,
1996; Chabrier \& M\'era, 1997).
The comparison with observations has already been presented in different
{\it Letters} (see references), and is examined in
detail in two companion papers, namely Baraffe et al. (1997) for metal-poor
globular cluster and halo field stars and in Allard et al. (1997a) for solar-like
abundances.
%***The analysis of the Brown Dwarf regime based on dusty atmosphere
%models is under progress.******

\medskip
Tables 2-7 are available by anonymous ftp: 
\par
\hskip 1cm ftp ftp.ens-lyon.fr \par
\hskip 1cm username: anonymous \par
\hskip 1cm ftp $>$ cd /pub/users/CRAL/ibaraffe \par
\hskip 1cm ftp $>$ get CB97\_models \par
\hskip 1cm ftp $>$ quit
\bigskip

\begin{acknowledgements} We are deeply endebted to our collaborators F. Allard and
P. Hauschildt, for providing various model atmospheres and for numerous
discussions. We are also grateful to D. Saumon
and D. Alexander for providing their zero-metallicity atmosphere
models and their grainless opacities, respectively, and to W. D\"appen for computing zero-metallicity MHD EOS
upon request. We also acknowledge useful conversations with A. Burrows in
particular and with the Tucson group in general.
The computation were done on the  CRAY C90 of the Centre d'Etudes Nucl\'eaire
de Grenoble.
\end{acknowledgements}

%\section*{References}

\vfil\eject

\begin{table}
\caption{ Bottom of the convective enveloppe $R_{CE}$ normalised to the
radius of the star R as a function of mass and metallicity, for an age
of 10 Gyrs.
Comparison is made with grey models based on the Krishna-Swamy prescription for the 0.4 $\mso$ and [M/H]=-1.5. 
The values 0.57 and 0.62 $\msol$ correspond to the eclipsing binary system YY-Gem (Leung and Schneider 1978)}
\begin{tabular}{lcccccc}\hline
\hline\noalign{\smallskip}
[M/H]  & $M/\mso$ & ${\rm log} L/\lsol$ & R ($10^{10}$ cm) & $R_{CE}$/R \\
\noalign{\smallskip}
\hline\noalign{\smallskip} 
 0
         & 0.4 & -1.71 & 2.640 &0.49 \\
         &0.5 & -1.45 & 3.285 &0.61 \\
         & 0.57& -1.25 & 3.806 &0.64 \\
         & 0.6 & -1.15 & 4.035 &0.65 \\
         &0.62 & -1.09& 4.189 &0.66 \\
%&&&&&\\
%0 $base$
%         & 0.4 & -1.80 & 2.68 &0.50 \\
%         & 0.5 & -1.53 & 3.31 &0.60 \\
%         &0.57 &-1.31 & 3.79 &0.64 \\
%         &0.6&-1.22 & 4.02 & 0.65\\
%         &.62&-1.15& 4.18 &0.66  \\
\hline
-1       &0.35 & -1.66&2.269& 0.43 \\
         &0.40 & -1.53 &2.566 & 0.58\\
         &0.50 & -1.18 & 3.335 & 0.69\\
         &0.60 & -0.782 & 4.013 & 0.73\\
\hline
-1.5     &0.4& -1.47 &2.517 & 0.58\\
         &0.5 & -1.12 & 3.252 & 0.69\\
         &0.60 & -0.70 & 3.904  & 0.75\\
&&&&&\\
-1.5 $Grey$  &0.4& -1.42 &2.455 & 0.54\\
\hline
-2       &0.4& -1.42 & 2.460 & 0.55\\
         &0.5 & -1.08 & 3.148 & 0.69\\
         &0.60 & -0.67 & 3.837  & 0.76\\
\hline
\end{tabular}
\end{table}

\vfill
\eject

\begin{table*}
\caption{ Properties of Very Low mass stars for [M/H]=0 and Y=0.275. Central
temperature $T_c$ is in K and density  $\rho_c$ in $gr \, cm^{-3}$. Abundances
of light elements are normalised to their initial abundance (see text).}
\begin{tabular}{lccccccccc}
\hline\noalign{\smallskip}
$M/\msol$ & age (Gyrs) & $\te$ &  log $L/\lsol$ & R ($10^{10}$ cm) & log $T_c$ &
log $\rho_c$ &  Li/Li$_0$ & Be/Be$_0$ & B/B$_0$ \\
\noalign{\smallskip}
\hline\noalign{\smallskip}
 0.075& 0.01&3006.&-2.048& 2.449&6.199&1.108&1.00000&1.00000&1.00000 \\ 
      & 0.10&2835.&-2.831& 1.118&6.459&2.125&0.35600&0.99800&1.00000 \\
      & 1.00&2211.&-3.695& 0.680&6.510&2.789&0.00000&0.03180&1.00000 \\
      &10.00&2002.&-3.929& 0.634&6.492&2.882&0.00000&0.00000&0.99667 \\
 0.080& 0.01&3025.&-2.010& 2.528&6.215&1.097&1.00000&1.00000&1.00000 \\ 
      & 0.10&2876.&-2.780& 1.152&6.482&2.117&0.17100&0.99400&1.00000 \\
      & 1.00&2374.&-3.536& 0.708&6.554&2.765&0.00000&0.00120&0.99333 \\
      &10.00&2314.&-3.605& 0.689&6.552&2.803&0.00000&0.00000&0.91667 \\
 0.090& 0.01&3059.&-1.939& 2.682&6.242&1.076&1.00000&1.00000&1.00000 \\ 
      & 0.10&2946.&-2.694& 1.213&6.522&2.106&0.02050&0.96100&1.00000 \\
      & 1.00&2645.&-3.261& 0.784&6.617&2.686&0.00000&0.00000&0.90333 \\
      &10.00&2642.&-3.265& 0.781&6.619&2.691&0.00000&0.00000&0.16200 \\
 0.100& 0.01&3090.&-1.874& 2.833&6.266&1.054&1.00000&1.00000&1.00000 \\ 
      & 0.10&3006.&-2.614& 1.278&6.555&2.087&0.00180&0.85200&1.00000 \\
      & 1.00&2811.&-3.070& 0.863&6.658&2.607&0.00000&0.00000&0.59333 \\
      &10.00&2814.&-3.069& 0.863&6.659&2.607&0.00000&0.00000&0.00038 \\
 0.110& 0.01&3112.&-1.821& 2.968&6.288&1.038&1.00000&1.00000&1.00000 \\ 
      & 0.10&3051.&-2.550& 1.334&6.585&2.075&0.00009&0.61000&1.00000 \\
      & 1.00&2919.&-2.932& 0.939&6.687&2.539&0.00000&0.00000&0.26533 \\
      &10.00&2922.&-2.929& 0.940&6.688&2.538&0.00000&0.00000&0.00000 \\
 0.150& 0.01&3186.&-1.636& 3.503&6.354&0.964&1.00000&1.00000&1.00000 \\ 
      & 0.10&3199.&-2.328& 1.567&6.669&2.009&0.00000&0.02640&0.99667 \\
      & 1.00&3149.&-2.581& 1.209&6.759&2.349&0.00000&0.00000&0.00293 \\
      &10.00&3153.&-2.577& 1.212&6.760&2.346&0.00000&0.00000&0.00000 \\
 0.200& 0.01&3251.&-1.464& 4.101&6.412&0.889&1.00000&1.00000&1.00000 \\ 
      & 0.10&3299.&-2.137& 1.835&6.738&1.934&0.00000&0.00009&0.91000 \\
      & 1.00&3290.&-2.316& 1.502&6.813&2.196&0.00000&0.00000&0.00000 \\
      &10.00&3295.&-2.302& 1.523&6.810&2.177&0.00000&0.00000&0.00000 \\
 0.300& 0.01&3345.&-1.232& 5.060&6.498&0.798&0.98700&1.00000&1.00000 \\ 
      & 0.10&3424.&-1.873& 2.310&6.818&1.831&0.00000&0.00000&0.79333 \\
      & 1.00&3437.&-1.984& 2.016&6.879&1.996&0.00000&0.00000&0.00000 \\
      &10.00&3436.&-1.945& 2.112&6.863&1.935&0.00000&0.00000&0.00000 \\
 0.350& 0.01&3396.&-1.138& 5.469&6.531&0.766&0.95300&1.00000&1.00000 \\ 
      & 0.10&3471.&-1.767& 2.540&6.836&1.821&0.00000&0.00000&0.97667 \\
      & 1.00&3478.&-1.857& 2.282&6.897&1.907&0.00000&0.00000&0.00847 \\
      &10.00&3475.&-1.822& 2.380&6.882&1.850&0.00000&0.00000&0.00000 \\
 0.400& 0.01&3451.&-1.050& 5.863&6.559&0.735&0.87600&1.00000&1.00000 \\ 
      & 0.10&3525.&-1.660& 2.786&6.853&1.827&0.00000&0.00016&0.99667 \\
      & 1.00&3522.&-1.727& 2.581&6.905&1.852&0.00000&0.00000&0.82667 \\
      &10.00&3524.&-1.707& 2.640&6.897&1.859&0.00000&0.00000&0.13533 \\
 0.500& 0.01&3555.&-0.903& 6.547&6.608&0.691&0.52200&0.99700&1.00000 \\ 
      & 0.10&3658.&-1.429& 3.374&6.898&1.875&0.00000&0.30000&1.00000 \\
      & 1.00&3649.&-1.478& 3.205&6.935&1.848&0.00000&0.09220&1.00000 \\
      &10.00&3654.&-1.454& 3.285&6.930&1.880&0.00000&0.00000&1.00000 \\
 0.600& 0.01&3645.&-0.785& 7.134&6.645&0.668&0.28900&0.99300&1.00000 \\ 
      & 0.10&3987.&-1.119& 4.058&6.979&1.935&0.00000&0.86400&1.00000 \\
      & 1.00&3883.&-1.195& 3.919&6.969&1.860&0.00000&0.84200&1.00000 \\
      &10.00&3914.&-1.156& 4.035&6.976&1.921&0.00000&0.70200&1.00000 \\
 0.700& 0.01&3719.&-0.681& 7.721&6.669&0.655&0.38200&0.99600&1.00000 \\ 
      & 0.10&4246.&-0.910& 4.551&7.027&1.880&0.00205&0.97400&1.00000 \\
      & 1.00&4209.&-0.909& 4.637&7.009&1.876&0.00105&0.97300&1.00000 \\
      &10.00&4284.&-0.841& 4.841&7.031&1.992&0.00000&0.96800&1.00000 \\
 0.800& 0.01&4091.&-0.539& 7.515&6.729&0.796&0.10400&0.98900&1.00000 \\
 & 0.10&4647.&-0.647& 5.140&7.059&1.876&0.00173&0.97000&1.00000 \\
 & 1.00&4647.&-0.632& 5.235&7.051&1.897&0.00157&0.97000&1.00000 \\
 &10.00&4762.&-0.515& 5.705&7.097&2.111&0.00085&0.96900&1.00000 \\
\hline
\end{tabular}
\end{table*}

\vfill
\eject

\begin{table*}
\caption{Same as Table 2 for [M/H]=-0.5 and Y=0.25}
\begin{tabular}{lccccccccc}
\hline\noalign{\smallskip}
$M/\msol$ & age (Gyrs) & $\te$ &  log $L/\lsol$ & R ($10^{10}$ cm) & log $T_c$ &
log $\rho_c$ &  Li/Li$_0$ & Be/Be$_0$ & B/B$_0$ \\
\noalign{\smallskip}
\hline\noalign{\smallskip}
 0.080& 0.01&3211.&-1.988& 2.299&6.238&1.218&1.00000&1.00000&1.00000 \\ 
      & 0.10&3023.&-2.759& 1.068&6.483&2.215&0.09320&0.99000&1.00000 \\
      & 1.00&2352.&-3.608& 0.664&6.508&2.849&0.00000&0.00772&1.00000 \\
      &10.00&2128.&-3.835& 0.625&6.485&2.928&0.00000&0.00000&0.99333 \\
 0.090& 0.01&3246.&-1.918& 2.441&6.266&1.196&1.00000&1.00000&1.00000 \\ 
      & 0.10&3102.&-2.666& 1.129&6.525&2.198&0.00708&0.93400&1.00000 \\
      & 1.00&2725.&-3.261& 0.738&6.594&2.764&0.00000&0.00000&0.95333 \\
      &10.00&2716.&-3.271& 0.734&6.595&2.772&0.00000&0.00000&0.45333 \\
 0.100& 0.01&3279.&-1.852& 2.580&6.290&1.173&1.00000&1.00000&1.00000 \\ 
      & 0.10&3160.&-2.593& 1.184&6.562&2.186&0.00026&0.73100&1.00000 \\
      & 1.00&2940.&-3.035& 0.822&6.644&2.671&0.00000&0.00000&0.67000 \\
      &10.00&2943.&-3.033& 0.823&6.646&2.670&0.00000&0.00000&0.00282 \\
 0.110& 0.01&3309.&-1.797& 2.700&6.313&1.158&1.00000&1.00000&1.00000 \\ 
      & 0.10&3211.&-2.522& 1.244&6.591&2.167&0.00001&0.44300&1.00000 \\
      & 1.00&3068.&-2.882& 0.900&6.677&2.595&0.00000&0.00000&0.29133 \\
      &10.00&3072.&-2.879& 0.901&6.679&2.593&0.00000&0.00000&0.00000 \\
 0.150& 0.01&3399.&-1.606& 3.187&6.380&1.085&1.00000&1.00000&1.00000 \\ 
      & 0.10&3353.&-2.310& 1.457&6.681&2.103&0.00000&0.00543&0.99000 \\
      & 1.00&3311.&-2.521& 1.172&6.754&2.390&0.00000&0.00000&0.00320 \\
      &10.00&3314.&-2.517& 1.175&6.756&2.386&0.00000&0.00000&0.00000 \\
 0.200& 0.01&3479.&-1.435& 3.704&6.442&1.020&0.99800&1.00000&1.00000 \\ 
      & 0.10&3467.&-2.115& 1.705&6.752&2.030&0.00000&0.00000&0.77333 \\
      & 1.00&3451.&-2.257& 1.462&6.809&2.230&0.00000&0.00000&0.00000 \\
      &10.00&3455.&-2.241& 1.485&6.806&2.211&0.00000&0.00000&0.00000 \\
 0.300& 0.01&3595.&-1.193& 4.586&6.526&0.924&0.94000&1.00000&1.00000 \\ 
      & 0.10&3640.&-1.827& 2.154&6.831&1.928&0.00000&0.00000&0.90000 \\
      & 1.00&3640.&-1.909& 1.960&6.877&2.032&0.00000&0.00000&0.00000 \\
      &10.00&3645.&-1.865& 2.058&6.860&1.969&0.00000&0.00000&0.00000 \\
 0.350& 0.01&3648.&-1.100& 4.956&6.560&0.892&0.81000&0.99900&1.00000 \\ 
      & 0.10&3703.&-1.708& 2.389&6.850&1.928&0.00000&0.00027&0.99667 \\
      & 1.00&3696.&-1.759& 2.260&6.888&1.923&0.00000&0.00000&0.19033 \\
      &10.00&3697.&-1.742& 2.305&6.883&1.895&0.00000&0.00000&0.00000 \\
 0.400& 0.01&3702.&-1.016& 5.298&6.589&0.865&0.58500&0.99800&1.00000 \\ 
      & 0.10&3777.&-1.582& 2.656&6.874&1.947&0.00000&0.16200&1.00000 \\
      & 1.00&3754.&-1.630& 2.541&6.903&1.905&0.00000&0.00027&0.99000 \\
      &10.00&3759.&-1.609& 2.598&6.899&1.919&0.00000&0.00000&0.94667 \\
 0.500& 0.01&3810.&-0.875& 5.887&6.633&0.841&0.44400&0.99700&1.00000 \\ 
      & 0.10&4033.&-1.269& 3.338&6.950&1.997&0.00011&0.95900&1.00000 \\
      & 1.00&3947.&-1.333& 3.237&6.941&1.901&0.00000&0.93700&1.00000 \\
      &10.00&3968.&-1.298& 3.333&6.944&1.954&0.00000&0.84900&1.00000 \\
 0.600& 0.01&3920.&-0.748& 6.433&6.662&0.847&0.67000&0.99900&1.00000 \\ 
      & 0.10&4357.&-1.008& 3.862&7.002&1.941&0.25400&0.99700&1.00000 \\
      & 1.00&4321.&-1.000& 3.960&6.986&1.914&0.22000&0.99700&1.00000 \\
      &10.00&4406.&-0.935& 4.109&7.004&2.020&0.11200&0.99600&1.00000 \\
 0.700& 0.01&4039.&-0.624& 6.990&6.688&0.871&0.86300&1.00000&1.00000 \\ 
      & 0.10&4804.&-0.704& 4.509&7.038&1.921&0.77100&1.00000&1.00000 \\
      & 1.00&4806.&-0.687& 4.593&7.032&1.929&0.76500&1.00000&1.00000 \\
      &10.00&4962.&-0.569& 4.933&7.075&2.136&0.74600&1.00000&1.00000 \\
 0.800& 0.01&4176.&-0.498& 7.560&6.715&0.916&0.95300&1.00000&1.00000 \\ 
      & 0.10&5217.&-0.432& 5.228&7.075&1.923&0.94200&1.00000&1.00000 \\
      & 1.00&5225.&-0.412& 5.334&7.076&1.941&0.94100&1.00000&1.00000 \\
      &10.00&5490.&-0.190& 6.239&7.164&2.359&0.94100&1.00000&1.00000 \\
\hline
\end{tabular}
\end{table*}

\vfill
\eject

\begin{table*}
\caption{Same as Table 2 for [M/H]=-1 and Y=0.25}
\begin{tabular}{lccccccccc}
\hline\noalign{\smallskip}
$M/\msol$ & age (Gyrs) & $\te$ &  log $L/\lsol$ & R ($10^{10}$ cm) & log $T_c$ &
log $\rho_c$ &  Li/Li$_0$ & Be/Be$_0$ & B/B$_0$ \\
\noalign{\smallskip}
\hline\noalign{\smallskip}
 0.083& 0.01&3421.&-1.946& 2.128&6.282&1.337&1.00000&1.00000&1.00000 \\ 
      & 0.10&3203.&-2.705& 1.013&6.511&2.303&0.00747&0.94700&1.00000 \\
      & 1.00&2527.&-3.504& 0.649&6.521&2.897&0.00000&0.00054&0.99667 \\
      &10.00&2359.&-3.660& 0.622&6.504&2.952&0.00000&0.00000&0.98333 \\
 0.085& 0.01&3429.&-1.930& 2.157&6.287&1.330&1.00000&1.00000&1.00000 \\ 
      & 0.10&3218.&-2.690& 1.020&6.521&2.304&0.00318&0.91700&1.00000 \\
      & 1.00&2636.&-3.411& 0.664&6.543&2.878&0.00000&0.00008&0.99333 \\
      &10.00&2550.&-3.492& 0.646&6.536&2.914&0.00000&0.00000&0.93667 \\
 0.090& 0.01&3449.&-1.897& 2.214&6.302&1.323&1.00000&1.00000&1.00000 \\ 
      & 0.10&3258.&-2.644& 1.050&6.542&2.294&0.00050&0.81300&1.00000 \\
      & 1.00&2848.&-3.222& 0.706&6.589&2.822&0.00000&0.00000&0.94667 \\
      &10.00&2840.&-3.231& 0.703&6.590&2.829&0.00000&0.00000&0.47000 \\
 0.100& 0.01&3486.&-1.832& 2.338&6.327&1.301&1.00000&1.00000&1.00000 \\ 
      & 0.10&3327.&-2.564& 1.104&6.579&2.277&0.00000&0.46400&1.00000 \\
      & 1.00&3103.&-2.970& 0.795&6.646&2.715&0.00000&0.00000&0.59000 \\
      &10.00&3107.&-2.967& 0.796&6.648&2.713&0.00000&0.00000&0.00079 \\
 0.110& 0.01&3518.&-1.771& 2.459&6.348&1.279&1.00000&1.00000&1.00000 \\ 
      & 0.10&3381.&-2.499& 1.153&6.612&2.266&0.00000&0.15900&1.00000 \\
      & 1.00&3240.&-2.812& 0.875&6.681&2.632&0.00000&0.00000&0.20800 \\
      &10.00&3245.&-2.808& 0.876&6.683&2.631&0.00000&0.00000&0.00000 \\
 0.150& 0.01&3617.&-1.581& 2.899&6.417&1.208&0.99900&1.00000&1.00000 \\ 
      & 0.10&3549.&-2.273& 1.356&6.704&2.197&0.00000&0.00025&0.95333 \\
      & 1.00&3500.&-2.445& 1.143&6.761&2.422&0.00000&0.00000&0.00111 \\
      &10.00&3505.&-2.439& 1.149&6.762&2.416&0.00000&0.00000&0.00000 \\
 0.200& 0.01&3701.&-1.406& 3.384&6.478&1.137&0.98600&1.00000&1.00000 \\ 
      & 0.10&3682.&-2.069& 1.593&6.776&2.118&0.00000&0.00000&0.46667 \\
      & 1.00&3666.&-2.173& 1.426&6.817&2.263&0.00000&0.00000&0.00000 \\
      &10.00&3674.&-2.150& 1.458&6.812&2.234&0.00000&0.00000&0.00000 \\
 0.300& 0.01&3831.&-1.159& 4.200&6.562&1.038&0.73100&0.99900&1.00000 \\ 
      & 0.10&3837.&-1.782& 2.042&6.855&1.990&0.00000&0.00000&0.75667 \\
      & 1.00&3834.&-1.836& 1.923&6.884&2.057&0.00000&0.00000&0.00000 \\
      &10.00&3841.&-1.787& 2.027&6.865&1.988&0.00000&0.00000&0.00000 \\
 0.350& 0.01&3881.&-1.070& 4.534&6.596&1.008&0.42500&0.99500&1.00000 \\ 
      & 0.10&3904.&-1.659& 2.273&6.874&1.993&0.00000&0.00074&0.99667 \\
      & 1.00&3891.&-1.683& 2.226&6.894&1.942&0.00000&0.00000&0.16667 \\
      &10.00&3893.&-1.665& 2.269&6.891&1.926&0.00000&0.00000&0.00000 \\
 0.400& 0.01&3930.&-0.989& 4.853&6.622&0.984&0.27700&0.99300&1.00000 \\ 
      & 0.10&3984.&-1.524& 2.549&6.904&2.012&0.00000&0.51500&1.00000 \\
      & 1.00&3950.&-1.553& 2.509&6.909&1.928&0.00000&0.00496&0.99667 \\
      &10.00&3959.&-1.530& 2.566&6.909&1.950&0.00000&0.00000&0.98333 \\
 0.500& 0.01&4035.&-0.848& 5.413&6.658&0.980&0.58200&0.99900&1.00000 \\ 
      & 0.10&4261.&-1.216& 3.179&6.973&2.002&0.08520&0.99400&1.00000 \\
      & 1.00&4190.&-1.231& 3.230&6.951&1.927&0.03000&0.99200&1.00000 \\
      &10.00&4235.&-1.185& 3.335&6.960&1.995&0.00014&0.98800&1.00000 \\
 0.600& 0.01&4169.&-0.708& 5.956&6.690&1.015&0.87700&1.00000&1.00000 \\ 
      & 0.10&4728.&-0.886& 3.775&7.011&1.955&0.81300&1.00000&1.00000 \\
      & 1.00&4711.&-0.873& 3.859&7.002&1.938&0.80200&1.00000&1.00000 \\
      &10.00&4867.&-0.782& 4.013&7.031&2.079&0.75400&1.00000&1.00000 \\
 0.700& 0.01&4377.&-0.543& 6.534&6.726&1.073&0.97100&1.00000&1.00000 \\ 
      & 0.10&5236.&-0.575& 4.402&7.050&1.939&0.96700&1.00000&1.00000 \\
      & 1.00&5236.&-0.559& 4.483&7.048&1.943&0.96700&1.00000&1.00000 \\
      &10.00&5478.&-0.392& 4.967&7.113&2.234&0.96700&1.00000&1.00000 \\
 0.800& 0.01&4663.&-0.349& 7.195&6.770&1.163&0.99400&1.00000&1.00000 \\ 
      & 0.10&5662.&-0.303& 5.150&7.089&1.944&0.99400&1.00000&1.00000 \\
      & 1.00&5719.&-0.269& 5.247&7.097&1.990&0.99400&1.00000&1.00000 \\
      &10.00&6055.& 0.077& 6.973&7.235&2.676&0.99400&1.00000&1.00000 \\
\hline
\end{tabular}
\end{table*}

\vfill
\eject
\begin{table*}
\caption{Same as Table 2 for [M/H]=-1.3 and Y=0.25}
\begin{tabular}{lccccccccc}
\hline\noalign{\smallskip}
$M/\msol$ & age (Gyrs) & $\te$ &  log $L/\lsol$ & R ($10^{10}$ cm) & log $T_c$ &
log $\rho_c$ &  Li/Li$_0$ & Be/Be$_0$ & B/B$_0$ \\
\noalign{\smallskip}
\hline\noalign{\smallskip}
 0.083& 0.01&3530.&-1.931& 2.031&6.299&1.397&1.00000&1.00000&1.00000 \\ 
      & 0.10&3283.&-2.699& 0.971&6.519&2.358&0.00181&0.90400&1.00000 \\
      & 1.00&2541.&-3.516& 0.632&6.510&2.930&0.00000&0.00049&0.99667 \\
      &10.00&2283.&-3.751& 0.597&6.480&3.003&0.00000&0.00000&0.99333 \\
 0.085& 0.01&3538.&-1.915& 2.061&6.304&1.390&1.00000&1.00000&1.00000 \\ 
      & 0.10&3304.&-2.677& 0.983&6.528&2.354&0.00082&0.86200&1.00000 \\
      & 1.00&2662.&-3.417& 0.646&6.534&2.913&0.00000&0.00006&0.99333 \\
      &10.00&2542.&-3.524& 0.626&6.523&2.954&0.00000&0.00000&0.96000 \\
 0.090& 0.01&3559.&-1.883& 2.113&6.319&1.384&1.00000&1.00000&1.00000 \\ 
      & 0.10&3347.&-2.631& 1.010&6.550&2.345&0.00009&0.70300&1.00000 \\
      & 1.00&2913.&-3.204& 0.690&6.585&2.854&0.00000&0.00000&0.94667 \\
      &10.00&2903.&-3.214& 0.686&6.586&2.862&0.00000&0.00000&0.49333 \\
 0.100& 0.01&3596.&-1.818& 2.231&6.345&1.362&1.00000&1.00000&1.00000 \\ 
      & 0.10&3419.&-2.551& 1.061&6.589&2.330&0.00000&0.30300&1.00000 \\
      & 1.00&3194.&-2.936& 0.781&6.647&2.738&0.00000&0.00000&0.53000 \\
      &10.00&3200.&-2.931& 0.782&6.649&2.736&0.00000&0.00000&0.00035 \\
 0.110& 0.01&3627.&-1.759& 2.347&6.366&1.340&1.00000&1.00000&1.00000 \\ 
      & 0.10&3479.&-2.480& 1.113&6.622&2.312&0.00000&0.08150&1.00000 \\
      & 1.00&3338.&-2.773& 0.862&6.684&2.653&0.00000&0.00000&0.16467 \\
      &10.00&3343.&-2.769& 0.863&6.686&2.651&0.00000&0.00000&0.00000 \\
 0.150& 0.01&3726.&-1.565& 2.781&6.434&1.262&0.99700&1.00000&1.00000 \\ 
      & 0.10&3658.&-2.251& 1.309&6.716&2.243&0.00000&0.00004&0.90667 \\
      & 1.00&3618.&-2.400& 1.127&6.765&2.441&0.00000&0.00000&0.00056 \\
      &10.00&3624.&-2.392& 1.134&6.766&2.433&0.00000&0.00000&0.00000 \\
 0.200& 0.01&3809.&-1.392& 3.247&6.495&1.191&0.96800&1.00000&1.00000 \\ 
      & 0.10&3785.&-2.048& 1.545&6.787&2.158&0.00000&0.00000&0.31667 \\
      & 1.00&3772.&-2.134& 1.409&6.821&2.279&0.00000&0.00000&0.00000 \\
      &10.00&3780.&-2.107& 1.447&6.815&2.244&0.00000&0.00000&0.00000 \\
 0.300& 0.01&3943.&-1.145& 4.026&6.580&1.094&0.53800&0.99700&1.00000 \\ 
      & 0.10&3940.&-1.758& 1.991&6.868&2.013&0.00000&0.00000&0.45667 \\
      & 1.00&3936.&-1.798& 1.906&6.887&2.068&0.00000&0.00000&0.00000 \\
      &10.00&3945.&-1.747& 2.011&6.868&1.998&0.00000&0.00000&0.00000 \\
 0.350& 0.01&3993.&-1.056& 4.353&6.613&1.061&0.23900&0.98700&1.00000 \\ 
      & 0.10&4005.&-1.639& 2.211&6.888&2.016&0.00000&0.00024&0.99667 \\
      & 1.00&3991.&-1.646& 2.208&6.897&1.949&0.00000&0.00000&0.04500 \\
      &10.00&3994.&-1.627& 2.255&6.890&1.919&0.00000&0.00000&0.00000 \\
 0.400& 0.01&4042.&-0.976& 4.654&6.637&1.042&0.19400&0.99100&1.00000 \\ 
      & 0.10&4093.&-1.498& 2.490&6.916&2.026&0.00000&0.55100&1.00000 \\
      & 1.00&4055.&-1.516& 2.483&6.913&1.934&0.00000&0.00262&0.99667 \\
      &10.00&4067.&-1.492& 2.537&6.914&1.960&0.00000&0.00000&0.97667 \\
 0.500& 0.01&4150.&-0.833& 5.205&6.672&1.049&0.64300&0.99900&1.00000 \\ 
      & 0.10&4373.&-1.190& 3.110&6.976&2.001&0.22000&0.99700&1.00000 \\
      & 1.00&4318.&-1.190& 3.188&6.956&1.934&0.11100&0.99600&1.00000 \\
      &10.00&4377.&-1.139& 3.291&6.968&2.008&0.00166&0.99300&1.00000 \\
 0.600& 0.01&4322.&-0.678& 5.736&6.709&1.100&0.92200&1.00000&1.00000 \\ 
      & 0.10&4897.&-0.841& 3.706&7.015&1.959&0.89900&1.00000&1.00000 \\
      & 1.00&4884.&-0.827& 3.783&7.007&1.946&0.89300&1.00000&1.00000 \\
      &10.00&5068.&-0.727& 3.944&7.042&2.100&0.86800&1.00000&1.00000 \\
 0.700& 0.01&4603.&-0.485& 6.322&6.752&1.180&0.98700&1.00000&1.00000 \\ 
      & 0.10&5409.&-0.531& 4.341&7.054&1.943&0.98600&1.00000&1.00000 \\
      & 1.00&5406.&-0.515& 4.423&7.052&1.943&0.98600&1.00000&1.00000 \\
      &10.00&5686.&-0.326& 4.970&7.128&2.273&0.98600&1.00000&1.00000 \\
 0.800& 0.01&4981.&-0.253& 7.045&6.809&1.299&0.99800&1.00000&1.00000 \\ 
      & 0.10&5843.&-0.258& 5.093&7.094&1.951&0.99800&1.00000&1.00000 \\
      & 1.00&5914.&-0.221& 5.187&7.105&2.005&0.99800&1.00000&1.00000 \\
      &10.00&6315.& 0.199& 7.373&7.268&2.841&0.99800&1.00000&1.00000 \\
\hline
\end{tabular}
\end{table*}

\begin{table*}
\caption{Same as Table 2 for [M/H]=-1.5 and Y=0.25}
\begin{tabular}{lccccccccc}
\hline\noalign{\smallskip}
$M/\msol$ & age (Gyrs) & $\te$ &  log $L/\lsol$ & R ($10^{10}$ cm) & log $T_c$ &
log $\rho_c$ &  Li/Li$_0$ & Be/Be$_0$ & B/B$_0$ \\
\noalign{\smallskip}
\hline\noalign{\smallskip}
 0.083& 0.01&3597.&-1.921& 1.979&6.309&1.432&1.00000&1.00000&1.00000 \\ 
      & 0.10&3336.&-2.688& 0.952&6.522&2.384&0.00092&0.87700&1.00000 \\
      & 1.00&2538.&-3.533& 0.622&6.502&2.952&0.00000&0.00044&0.99667 \\
      &10.00&2194.&-3.844& 0.582&6.459&3.039&0.00000&0.00000&0.99667 \\
 0.085& 0.01&3606.&-1.910& 1.997&6.316&1.431&1.00000&1.00000&1.00000 \\ 
      & 0.10&3354.&-2.672& 0.959&6.532&2.385&0.00030&0.81300&1.00000 \\
      & 1.00&2669.&-3.427& 0.635&6.528&2.935&0.00000&0.00005&0.99333 \\
      &10.00&2519.&-3.559& 0.613&6.512&2.982&0.00000&0.00000&0.97333 \\
 0.090& 0.01&3626.&-1.874& 2.058&6.329&1.418&1.00000&1.00000&1.00000 \\ 
      & 0.10&3399.&-2.625& 0.985&6.555&2.378&0.00003&0.61600&1.00000 \\
      & 1.00&2948.&-3.196& 0.679&6.583&2.874&0.00000&0.00000&0.94667 \\
      &10.00&2938.&-3.207& 0.675&6.583&2.882&0.00000&0.00000&0.51333 \\
 0.100& 0.01&3662.&-1.805& 2.184&6.353&1.390&1.00000&1.00000&1.00000 \\ 
      & 0.10&3475.&-2.545& 1.035&6.595&2.363&0.00000&0.21400&1.00000 \\
      & 1.00&3249.&-2.915& 0.773&6.648&2.753&0.00000&0.00000&0.51000 \\
      &10.00&3255.&-2.910& 0.774&6.650&2.750&0.00000&0.00000&0.00022 \\
 0.110& 0.01&3695.&-1.750& 2.286&6.377&1.375&1.00000&1.00000&1.00000 \\ 
      & 0.10&3542.&-2.470& 1.085&6.629&2.346&0.00000&0.04450&0.99667 \\
      & 1.00&3399.&-2.750& 0.854&6.685&2.665&0.00000&0.00000&0.13833 \\
      &10.00&3405.&-2.745& 0.856&6.687&2.663&0.00000&0.00000&0.00000 \\
 0.150& 0.01&3795.&-1.556& 2.707&6.444&1.297&0.99500&1.00000&1.00000 \\ 
      & 0.10&3722.&-2.239& 1.282&6.723&2.271&0.00000&0.00001&0.86000 \\
      & 1.00&3685.&-2.375& 1.119&6.767&2.451&0.00000&0.00000&0.00036 \\
      &10.00&3691.&-2.366& 1.126&6.768&2.443&0.00000&0.00000&0.00000 \\
 0.200& 0.01&3880.&-1.383& 3.162&6.505&1.226&0.94400&1.00000&1.00000 \\ 
      & 0.10&3848.&-2.033& 1.522&6.793&2.178&0.00000&0.00000&0.24867 \\
      & 1.00&3836.&-2.111& 1.400&6.824&2.288&0.00000&0.00000&0.00000 \\
      &10.00&3845.&-2.082& 1.441&6.816&2.250&0.00000&0.00000&0.00000 \\
 0.300& 0.01&4013.&-1.136& 3.930&6.589&1.125&0.41900&0.99500&1.00000 \\ 
      & 0.10&4001.&-1.745& 1.960&6.875&2.032&0.00000&0.00000&0.16833 \\
      & 1.00&3999.&-1.775& 1.896&6.889&2.075&0.00000&0.00000&0.00000 \\
      &10.00&4008.&-1.724& 2.002&6.870&2.004&0.00000&0.00000&0.00000 \\
 0.350& 0.01&4064.&-1.046& 4.245&6.623&1.094&0.15400&0.97900&1.00000 \\ 
      & 0.10&4068.&-1.620& 2.188&6.893&2.022&0.00000&0.00009&0.99667 \\
      & 1.00&4058.&-1.642& 2.145&6.907&1.989&0.00000&0.00000&0.01817 \\
      &10.00&4059.&-1.602& 2.247&6.891&1.924&0.00000&0.00000&0.00000 \\
 0.400& 0.01&4112.&-0.968& 4.540&6.647&1.077&0.14400&0.98900&1.00000 \\ 
      & 0.10&4157.&-1.485& 2.451&6.922&2.032&0.00000&0.49700&1.00000 \\
      & 1.00&4122.&-1.495& 2.463&6.915&1.937&0.00000&0.00088&0.99000 \\
      &10.00&4134.&-1.471& 2.517&6.917&1.964&0.00000&0.00000&0.96333 \\
 0.500& 0.01&4232.&-0.820& 5.084&6.682&1.087&0.63700&0.99900&1.00000 \\ 
      & 0.10&4435.&-1.178& 3.065&6.976&1.995&0.25500&0.99700&1.00000 \\
      & 1.00&4396.&-1.169& 3.150&6.958&1.937&0.11100&0.99600&1.00000 \\
      &10.00&4462.&-1.116& 3.252&6.971&2.013&0.00107&0.99200&1.00000 \\
 0.600& 0.01&4427.&-0.657& 5.603&6.721&1.148&0.93500&1.00000&1.00000 \\ 
      & 0.10&4977.&-0.821& 3.669&7.016&1.957&0.92200&1.00000&1.00000 \\
      & 1.00&4970.&-0.807& 3.738&7.009&1.946&0.91700&1.00000&1.00000 \\
      &10.00&5167.&-0.702& 3.904&7.047&2.108&0.90200&1.00000&1.00000 \\
 0.700& 0.01&4760.&-0.444& 6.194&6.769&1.239&0.99200&1.00000&1.00000 \\ 
      & 0.10&5497.&-0.510& 4.303&7.057&1.948&0.99200&1.00000&1.00000 \\
      & 1.00&5494.&-0.494& 4.388&7.055&1.947&0.99200&1.00000&1.00000 \\
      &10.00&5792.&-0.296& 4.959&7.135&2.292&0.99200&1.00000&1.00000 \\
 0.800& 0.01&5167.&-0.195& 7.004&6.834&1.376&0.99900&1.00000&1.00000 \\ 
      & 0.10&5933.&-0.238& 5.054&7.096&1.951&0.99900&1.00000&1.00000 \\
      & 1.00&6021.&-0.198& 5.138&7.109&2.015&0.99900&1.00000&1.00000 \\
      &10.00&6483.& 0.266& 7.557&7.296&2.926&0.99900&1.00000&1.00000 \\
\hline
\end{tabular}
\end{table*}

\vfill
\eject

\begin{table*}
\caption{Same as Table 2 for [M/H]=-2 and Y=0.25}
\begin{tabular}{lccccccccc}
\hline\noalign{\smallskip}
$M/\msol$ & age (Gyrs) & $\te$ &  log $L/\lsol$ & R ($10^{10}$ cm) & log $T_c$ &
log $\rho_c$ &  Li/Li$_0$ & Be/Be$_0$ & B/B$_0$ \\
 \noalign{\smallskip}
\hline\noalign{\smallskip}
 0.083& 0.01&3750.&-1.900& 1.867&6.330&1.507&1.00000&1.00000&1.00000 \\ 
      & 0.10&3455.&-2.676& 0.900&6.531&2.459&0.00009&0.76300&1.00000 \\
      & 1.00&2519.&-3.578& 0.599&6.480&3.000&0.00000&0.00045&0.99667 \\
      &10.00&1779.&-4.274& 0.539&6.371&3.135&0.00000&0.00000&0.99667 \\
 0.085& 0.01&3758.&-1.889& 1.883&6.337&1.507&1.00000&1.00000&1.00000 \\ 
      & 0.10&3479.&-2.654& 0.910&6.542&2.455&0.00003&0.67400&1.00000 \\
      & 1.00&2667.&-3.461& 0.611&6.509&2.984&0.00000&0.00004&0.99333 \\
      &10.00&2342.&-3.741& 0.575&6.468&3.063&0.00000&0.00000&0.99000 \\
 0.090& 0.01&3779.&-1.852& 1.941&6.351&1.495&1.00000&1.00000&1.00000 \\ 
      & 0.10&3532.&-2.605& 0.934&6.566&2.449&0.00000&0.41500&1.00000 \\
      & 1.00&3022.&-3.186& 0.654&6.574&2.924&0.00000&0.00000&0.95000 \\
      &10.00&3005.&-3.202& 0.649&6.574&2.934&0.00000&0.00000&0.60333 \\
 0.100& 0.01&3816.&-1.784& 2.060&6.375&1.466&1.00000&1.00000&1.00000 \\ 
      & 0.10&3623.&-2.520& 0.979&6.608&2.437&0.00000&0.08090&1.00000 \\
      & 1.00&3387.&-2.866& 0.753&6.649&2.788&0.00000&0.00000&0.43333 \\
      &10.00&3395.&-2.860& 0.754&6.652&2.785&0.00000&0.00000&0.00006 \\
 0.110& 0.01&3846.&-1.727& 2.167&6.397&1.444&0.99900&1.00000&1.00000 \\ 
      & 0.10&3696.&-2.441& 1.031&6.642&2.414&0.00000&0.01050&0.99667 \\
      & 1.00&3556.&-2.692& 0.834&6.689&2.696&0.00000&0.00000&0.09867 \\
      &10.00&3565.&-2.686& 0.836&6.691&2.693&0.00000&0.00000&0.00000 \\
 0.150& 0.01&3947.&-1.535& 2.566&6.466&1.367&0.98400&1.00000&1.00000 \\ 
      & 0.10&3873.&-2.207& 1.229&6.736&2.326&0.00000&0.00000&0.72333 \\
      & 1.00&3845.&-2.317& 1.098&6.773&2.475&0.00000&0.00000&0.00013 \\
      &10.00&3852.&-2.306& 1.109&6.773&2.463&0.00000&0.00000&0.00000 \\
 0.200& 0.01&4036.&-1.362& 2.993&6.528&1.298&0.84500&1.00000&1.00000 \\ 
      & 0.10&4002.&-1.997& 1.466&6.806&2.227&0.00000&0.00000&0.11800 \\
      & 1.00&3995.&-2.054& 1.377&6.829&2.309&0.00000&0.00000&0.00000 \\
      &10.00&4004.&-2.020& 1.427&6.820&2.263&0.00000&0.00000&0.00000 \\
 0.300& 0.01&4189.&-1.114& 3.696&6.615&1.206&0.15700&0.97900&1.00000 \\ 
      & 0.10&4162.&-1.702& 1.904&6.887&2.069&0.00000&0.00000&0.00843 \\
      & 1.00&4161.&-1.717& 1.872&6.894&2.092&0.00000&0.00000&0.00000 \\
      &10.00&4172.&-1.665& 1.978&6.875&2.020&0.00000&0.00000&0.00000 \\
 0.350& 0.01&4252.&-1.021& 3.996&6.649&1.173&0.03360&0.92400&1.00000 \\ 
      & 0.10&4231.&-1.587& 2.104&6.911&2.036&0.00000&0.00000&0.92333 \\
      & 1.00&4226.&-1.589& 2.104&6.915&2.009&0.00000&0.00000&0.00000 \\
      &10.00&4231.&-1.541& 2.218&6.897&1.941&0.00000&0.00000&0.00000 \\
 0.400& 0.01&4306.&-0.943& 4.258&6.673&1.162&0.02590&0.96200&1.00000 \\ 
      & 0.10&4315.&-1.456& 2.349&6.935&2.033&0.00000&0.08290&1.00000 \\
      & 1.00&4289.&-1.444& 2.411&6.921&1.939&0.00000&0.00000&0.92333 \\
      &10.00&4304.&-1.421& 2.460&6.925&1.972&0.00000&0.00000&0.69667 \\
 0.500& 0.01&4434.&-0.791& 4.789&6.707&1.172&0.38600&0.99700&1.00000 \\ 
      & 0.10&4589.&-1.148& 2.963&6.979&1.996&0.06000&0.99300&1.00000 \\
      & 1.00&4560.&-1.136& 3.042&6.963&1.941&0.00305&0.98400&1.00000 \\
      &10.00&4623.&-1.083& 3.148&6.977&2.021&0.00000&0.95600&1.00000 \\
 0.600& 0.01&4643.&-0.621& 5.308&6.747&1.234&0.89400&1.00000&1.00000 \\ 
      & 0.10&5093.&-0.799& 3.596&7.017&1.958&0.87700&1.00000&1.00000 \\
      & 1.00&5094.&-0.784& 3.655&7.012&1.951&0.87100&1.00000&1.00000 \\
      &10.00&5292.&-0.676& 3.837&7.052&2.116&0.85200&1.00000&1.00000 \\
 0.700& 0.01&5000.&-0.401& 5.899&6.800&1.335&0.98900&1.00000&1.00000 \\ 
      & 0.10&5596.&-0.489& 4.254&7.059&1.950&0.98900&1.00000&1.00000 \\
      & 1.00&5591.&-0.473& 4.340&7.057&1.946&0.98900&1.00000&1.00000 \\
      &10.00&5914.&-0.265& 4.932&7.143&2.312&0.98900&1.00000&1.00000 \\
 0.800& 0.01&5419.&-0.123& 6.917&6.871&1.485&0.99900&1.00000&1.00000 \\ 
      & 0.10&6045.&-0.216& 4.991&7.099&1.956&0.99900&1.00000&1.00000 \\
      & 1.00&6121.&-0.180& 5.078&7.111&2.007&0.99900&1.00000&1.00000 \\
      &10.00&6688.& 0.334& 7.682&7.314&3.021&0.99900&1.00000&1.00000 \\
\hline
\end{tabular}
\end{table*}

\end{document}